\documentclass[10pt, leqno]{amsart}
\usepackage{natbib}
\usepackage{amsmath}
\usepackage{graphicx,psfrag,epsf}
\usepackage{enumerate}
\usepackage{amsfonts}
\usepackage{algorithm}
\usepackage{algpseudocode}
\usepackage{dsfont}
\usepackage{url} 
\usepackage{hyperref}
\hypersetup{colorlinks,linkcolor={black},citecolor={black},urlcolor={black}} 
\usepackage{indentfirst,csquotes}

\usepackage{amssymb,amsthm,amsmath}
\usepackage{xcolor,paralist,hyperref,titlesec,fancyhdr,etoolbox}

\newtheorem{property}{Property}

\newcommand{\betabold}{\text{\boldmath$\alpha$}}
\newcommand{\R}{{\mathds R}}
\newcommand{\E}{{\mathds E}}
\renewcommand{\P}{\mathds{P}} 
\newcommand{\X}{{\bf X}}
\newcommand{\x}{{\bf x}}
\newcommand{\Y}{{\bf Y}}
\newcommand{\y}{{\bf y}}
\newcommand{\B}{{\bf B}}
\newcommand{\kATE}{SATE$_k$}

\newcommand{\lt}{{\bf l}}
\newcommand{\Norm}{\mathcal{N}}
\newcommand{\iv}{\mathbb{I}}
\DeclareRobustCommand{\rchi}{{\mathpalette\irchi\relax}}
\newcommand{\irchi}[2]{\raisebox{\depth}{$#1\chi$}} 
\newcommand{\PM}{\textsc{pm}$_{2.5}$ }
\newcommand{\PMns}{\textsc{pm}$_{2.5}$}

\newcommand{\review}[1]{{\color{black} #1}}

\titleformat{\section}[display]{\normalfont\huge\bfseries\centering}{\centering\chaptertitlename\thechapter}{10pt}{\Large}
\titlespacing*{\section}{0pt}{0ex}{0ex}

\hypersetup{ colorlinks=true, linkcolor=black, filecolor=black, urlcolor=black }

\usepackage{lipsum}

\begin{document}
\title{Multivariate Causal Effects: a Bayesian Causal Regression Factor Model} 
\author[]{
Dafne Zorzetto$^{1,*}$, 
Jenna Landy$^{2}$, Corwin Zigler$^{3}$, Giovanni Parmigiani$^{2,4}$, and
Roberta De Vito$^{1,3}$
}

\maketitle

\let\thefootnote\relax
\footnotetext{
$^*$ \texttt{dafne$\_$zorzetto@brown.edu}\\
$^{1}$ Data Science Institute, Brown University, Providence, Rhode Island, U.S.A. \\
$^{2}$ Department of Biostatistics, Harvard University, Cambridge, Massachusetts, U.S.A.\\
$^{3}$ Department of Biostatistics, Brown University, Providence, Rhode Island, U.S.A. \\
$^{4}$ Department of Data Science, Dana Farber Cancer Institute, Boston, Massachusetts, U.S.A.
} 

\begin{abstract}
The impact of wildfire smoke on air quality is a growing concern, contributing to air pollution through a complex mixture of chemical species with important implications for public health. Although previous studies have focused mainly on its association with total particulate matter (\PMns), the causal relationship between wildfire smoke and the chemical composition of \PM remains largely unexplored. Exposure to these chemical mixtures plays a critical role in shaping public health, but capturing their relationships requires advanced statistical methods capable of modeling the complex dependencies among chemical species. 
To fill this gap, we propose a Bayesian causal regression factor model that estimates the multivariate causal effects of wildfire smoke on the concentration of 27 chemical species in \PM across the United States. 
Our approach introduces two key innovations: (i) a causal inference framework for multivariate potential outcomes, and (ii) a novel Bayesian factor model that employs a probit stick-breaking process as prior for treatment-specific factor scores. By focusing on factor scores, our method addresses the missing data challenge common to causal inference and enables a flexible, data-driven characterization of the latent factor structure, which is crucial to capture the complex correlation between multivariate outcomes. 
Through Monte Carlo simulations, we show the accuracy of the model in estimating the causal effects in multivariate outcomes and characterizing the treatment-specific latent structure. Finally, we apply our method to US air quality data, estimating the causal effect of wildfire smoke on 27 chemical species in \PMns, providing a deeper understanding of their interdependencies.
\end{abstract} 

\bigskip


\noindent%
\vfill

\bigskip


\section*{1. Introduction}
\label{sec:1_intro}


In observational studies, a single cause or exposure can affect multiple outcomes that are often inherently strongly correlated. For instance, in environmental health, air pollution may simultaneously influence several diseases or mortality causes. In genomics, a cancer treatment can alter mutational signature profiles. Similarly, as analyzed in this paper, wildfire smoke represents a specific exposure that can change the chemical composition of ambient air pollutants.

Wildfires have emerged as a critical environmental and public health concern, significantly deteriorating air quality through the emission of smoke and associated pollutants \citep{childs2022daily,krasovich2025influence}. In the United States, wildfire smoke exposure has increased significantly, with the burned area quadrupling over the past four decades, a trend expected to continue as climate conditions become increasingly favorable to fire activity. Wildfire smoke can rise high into the atmosphere and travel long distances, affecting air pollution levels and health risks far from the source \citep{burke2021changing,o2021estimated}. As a result, wildfire pollution has received considerable attention in environmental health research \citep{burke2021changing, krasovich2025influence}.

A growing body of literature has examined the effects of wildfire smoke on particulate matter (\PMns), highlighting the need to quantify the proportion of \PM attributable to wildfires \citep{burke2021changing, childs2022daily}, given the well-documented adverse health effects of elevated \PM exposure \citep{carone2020pursuit,epa2022}. However, wildfire smoke is a complex mixture of pollutants that undergoes physical and chemical transformations during atmospheric transport.  Understanding how wildfire smoke influences the concentration levels of various chemical species in \PMns is therefore a pressing challenge.

While previous studies have focused on the impact of wildfire smoke on individual chemical components ---such as phosphorus and nitrogen \citep{spencer1991phosphorus}, lead \citep{odigie2014trace}, bromine and other metals \citep{li2023large}--- the recent work of \citet{krasovich2025influence} has provided a more comprehensive analysis by quantifying the contribution of wildfire smoke to the concentration of 27 different chemical species in \PM using a standard linear regression model. Their analysis focuses primarily on accounting for temporal variation and estimating the contribution of wildfire smoke to each chemical species in \PMns. However, it does not jointly model chemical species, limiting information sharing across species, and does not adopt a causal inference perspective, restricting its ability to disentangle the causal effect of wildfire smoke.

A causal inference framework is crucially needed to rigorously assess and quantify the effect of wildfire smoke on changes in chemical concentration levels, providing a deeper understanding of the associations highlighted in previous studies. A Bayesian perspective is particularly well suited to this setting, as it provides coherent uncertainty quantification for causal estimands, allows prior information to be incorporated in a principled way, and offers a flexible modeling framework for complex data structures \citep{li2022bayesian}.
Moreover, incorporating the high correlation structure among the different chemical species is critical to capturing deeper insights into the composition of smoke and its broader environmental effects.


To address these limitations, we propose a novel Bayesian factor model for causal inference, which leverages infinite mixture distributions as priors for the treatment-specific factor scores. The key novelty of our project lies in two main aspects. First, we define a causal inference framework within a Bayesian factor analysis model, to account for the high correlation between multivariate potential outcomes, such as the chemical species in \PMns. Second, we introduce a novel prior on the factor scores using a Dirichlet process \citep{quintana2020dependent} with a probit linear model to capture treatment-specific heterogeneity in a data-driven way. 

Traditional Bayesian factor models have focused primarily on priors for factor loadings \citep[e.g.,][]{bhattacharya2011sparse, schiavon2022generalized}, typically assuming standard normal distributions for factor scores. Recent works \citep{zorzetto2024sparse, bortolato2024adaptive} have begun exploring priors on factor scores, since they define the subject-specific contribution of each factor to the variation, making them essential in many applications. 
However, introducing a causal inference framework in
Bayesian factor models poses new challenges due to the fundamental missing data problem of causal inference \citep{rubin1980randomization}.  Our proposed prior addresses this by allowing a covariate-dependent structure in treatment-specific scores, enabling better estimation of latent confounding patterns. In particular, our prior leverages the flexibility of the Dependent Dirichlet Process (DDP) \citep{quintana2020dependent}, building on a growing body of work in causal inference that employs DDP priors \citep[e.g.,][]{roy2024bayesian, hu2023bayesian, zorzetto2024confounder}, though primarily in the context of scalar outcomes.

A key strength of this model is its flexibility and broad applicability to a wide range of real-world datasets where a treatment or exposure affects multivariate outcomes composed of correlated components. Beyond the wildfire application, the model can be used, for example, to estimate the effects of \PM exposure on cause-specific mortality or to investigate how cancer drugs affect transcriptional profiles.

The paper is organized as follows.  Section 2 introduces the causal inference setup, including the estimands of interest, causal assumptions, and identification strategy. Section 3 introduces our Bayesian factor model for causal inference, highlighting the innovation in our prior specification for factor scores and its critical role in this context. In Section 4, we assess the performance of the model performance through simulations. In Section 5, we apply our approach to estimate the causal effect of the presence of wildfire smoke on the concentration of 27 chemical species in \PM across the United States. 
We conclude the paper with a discussion and future directions in Section 6.

\bigskip

\section*{2. Causal setup}
\label{sec:2_setup}
\bigskip \section*{\small Notation and Overview}

Let $i$ denote the study unit, with $i = 1, \dots, n$. For each unit, let $T_i \in \{0,1\}$  the binary treatment random variable with observed value $t_i$, let $\X_i =(X_{1i}, \dots, X_{pi})^T$ be the vector of $p$ observed covariates (confounders), which may be either continuous or discrete, and let $\mathbf{Y}_i \in \mathbb{R}^q$ be the multivariate outcome of dimension $q$. The outcome components are potentially highly correlated, a structure acknowledged by our Bayesian causal factor model. Throughout the paper, bold letters denote vectors. 

Following the definition of the Rubin Causal Model \citep{rubin1974estimating}, we make the Stable Unit Treatment Value Assumption \citep[SUTVA;][]{rubin1980randomization}, which is a combination of two assumptions: no interference among the units, i.e. the potential outcome values from unit $i$ do not depend on the treatment applied to other units, and consistency, i.e. each observed outcome is the individual's potential outcome under their treatment. Under SUTVA, we define the potential outcomes, for unit $i$, as $\{\Y_{i}(0), \Y_i(1)\}$, where $\Y_{i}(t)=(Y_{i1}(t), \dots, Y_{iq}(t))^T$ for $t\in \{0,1\}$.
Specifically, $\Y_{i}(0) \in \R^q$ is the $q$-variate outcome when the unit $i$ is assigned to the control group and $\Y_{i}(1) \in \R^q$ is the $q$-variate outcome when the unit $i$ is assigned to the treatment group. Under consistency, therefore the observed outcome can be written as
$
\Y_i = (1-T_i)\, \Y_i(0) + T_i \, \Y_i(1).
$
In practice, for $i=1,\dots,n$, we observe only $\y_i \in \R^q$, the realization of the random variable $\Y_i=(Y_{i1}, \dots ,Y_{iq})$. 
Conversely, we do not observe the missing outcome $\Y_i^{mis} \in \R^q$ defined as  $\Y_i^{mis}= T_i \, \Y_i(0) + (1-T_i) \, \Y_i(1)$. 

\begin{figure}[h!]
\centerline{
    \includegraphics[width=0.7\textwidth]{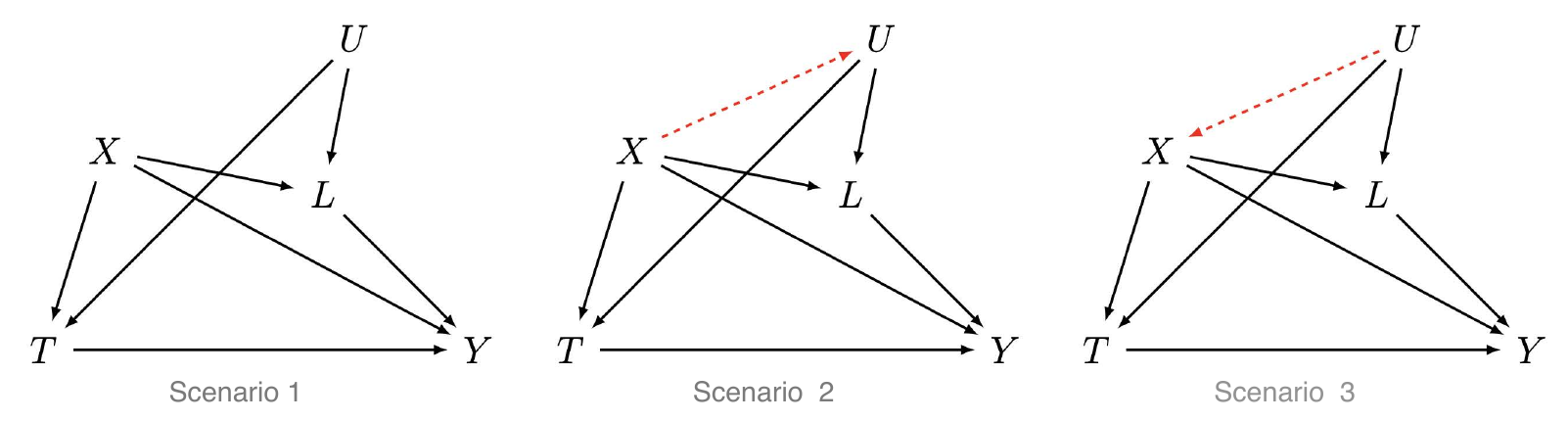}
}
\caption{Graphical representation of the causal pathway assumed in this paper, with treatment $T$, measured confounders $X$, unmeasured confounder $U$, latent factors $L$, and multivariate outcome $Y$. The three scenarios consider different relationships for unmeasured confounder $U$: (1) confounding only through an effect on latent factors $L$, (2) confounding as an effect of measured confounders $X$ on $U$, (3) confounding because $U$ is a cause of measured confounders $X$. The simulation study investigates each of these scenarios. \label{fig:dags}}
\end{figure}

To concisely define the correlated nature of the multidimensional outcome, for each unit $i$, 
we assume an unmeasured variable ${U}_i$ that affects the treatment $T_i$ and the potential outcome $\{\Y_{i}(0), \Y_i(1)\}$. 
Under the assumption of our proposed Bayesian causal factor model, formally introduced in Section \ref{sec:model}, we model both factual and counterfactual latent factors $\{{\bf L}_{i0}, {\bf L}_{i1}\}$ that capture the indirect effect of $U_i$ on the potential outcomes $\{\Y_{i}(0), \Y_i(1)\}$, as illustrated in Figure~\ref{fig:dags}
Figure~\ref{fig:dags}
shows the three versions of the resulting conditional independence structure considered in our model. 



In our environmental data application, the latent factors $L_t$, identified in our factor model, explain the observed correlation structure among chemical species. While the latent variables denoted by $U$ may represent specific pollution sources such as traffic, metalworking industries, energy production, or soil dust \citep[see e.g.][]{park2014assessment}, which are often impractical to measure comprehensively, but may determine the latent correlation structure.

\bigskip \section*{\small Causal Estimands and Identifying Assumptions}
\label{subsec:estimands}

Our goal is to estimate the vector of causal effects of the treatment on each component of the multivariate outcome. 
Therefore, we define the quantities of interest as the $q$-dimensional \textit{Sample Average Treatment Effects} (SATE) 
\begin{gather}
    \mbox{\textbf{SATE}} =\E[\Y(1)-\Y(0)], \notag 
\end{gather}
where the average is taken across the units $\{1, \dots, n\}$ 
and {\bf SATE} $\in \mathbb{R}^q$.

As common in causal inference, these causal estimands cannot be estimated directly from observed data due to the missing potential outcomes. To address this, we adopt the following assumptions and leverage our proposed factor model, specifically the relationship of treatment-specific factor scores ---i.e., $\lt_{0}$ and $\lt_{1}$---with measured and unmeasured variables. The first assumption is standard in the causal inference literature, the second is common in concept but adapted to our setting, while the third is newly defined for the context of the assumed latent factor structure.

\vspace{0.25cm}
\noindent {\em Positivity.} Each unit $i\in \{1, \dots, n\}$ has a non-zero probability of receiving each treatment level, such that $
0<\Pr(T_i=1 \mid \X_i=\x_i)<1.$

\vspace{0.25cm}
\noindent {\em Conditional ignorability.} The assignment of treatment is random in each group of units characterized by observed covariates $\x$ and latent factors scores $\lt_{0}$ and $\lt_{1}$, such that 
$
\{\Y_i(1), \Y_i(0)\} \perp T_i \mid \{ \X_i = \x_i, {\bf L}_{i0}=\lt_{i0}, {\bf L}_{i1}=\lt_{i1}\},
$ for each unit $i\in \{1, \dots, n\}$.

\vspace{0.25cm}
\noindent {\em Indirectness of unmeasured confounding effects.}
The unmeasured variable $U$ affects the potential outcomes only indirectly, through the latent factor and the observed confounders, such as
$
\P(\Y(1), \Y(0) \mid \X, {\bf L}_{0}, {\bf L}_{1}, U)=\P(\Y(1), \Y(0) \mid \X, {\bf L}_{0}, {\bf L}_{1}).
$
This assumption can be visualized in Figure \ref{fig:dags}
as the absence of a direct arrow between the $U$ and $\Y$.

\bigskip \section*{\small Identification}

Under the three assumptions of Section \ref{subsec:estimands}, the causal estimands (i.e., functions of potential outcomes) can be expressed as statistical estimands (i.e., functions of random variables) and can then be estimated using observed data.
\begin{property}
\label{prep:ATEs} If the three assumptions of Section~\ref{subsec:estimands} hold, the statistical estimand for SATEs is: 
\begin{align*}
    & \E[\Y(1) - \Y(0)] 
    = \int_{\mathcal{X}} \int_{\mathbb{R}^{J_0}}\int_{\mathbb{R}^{J_1}} \bigg(\E[{\bf Y} \mid T=1, \X=\x, {\bf L}_{0}=\lt_{0}, {\bf L}_{1}=\lt_{1}]\\
     & \quad \quad - \E[{\bf Y} \mid T=0, \X=\x, {\bf L}_{0}=\lt_{0}, {\bf L}_{1}=\lt_{1}]\bigg) 
     \times \P({\bf L}_{0}=\lt_{0}, {\bf L}_{1}=\lt_{1} \mid \X=\x)\P(\X=\x) d \lt_0, d \lt_1 d_\x;
\end{align*}
where the inner expectation $\E[{\bf Y} \mid T=t, \X=\x, {\bf L}_{0}=\lt_{0}, {\bf L}_{1}=\lt_{1}]$, for each treatment level $t \in \{0,1\}$, will be estimated using our proposed factor model \eqref{eq:model} and the joint probability of the latent factor scores $\P({\bf L}_{0}, {\bf L}_{1} \mid \X)$ modeled via prior distribution specified in  \eqref{eq:prior_factors}.
\end{property}


Property \ref{prep:ATEs} underlines the pivotal role of the probability distribution of treatment-specific factor scores and thus the necessity of defining a flexible prior, such as the one we propose in \eqref{eq:prior_factors}. Moreover, the inner expectation $\E[{\bf Y} \mid T=t, \X=\x, {\bf L}_{0}=\lt_{0}, {\bf L}_{1}=\lt_{1}]$ can be rewritten with an additional integration over the unmeasured variable $U$, as follows:
\begin{align}
    &\E[{\bf Y} \mid t, x, l_0, l_1] 
    = \int_\mathcal{U} \E[{\bf Y} \mid t, x, l_0, l_1, u]\P(u \mid t, x, l_0, l_1) d u \notag\\
    & \quad\quad\quad\quad\quad\quad\quad\; = \int_\mathcal{U} \frac{\E[{\bf Y} \mid t, x, l_0, l_1]\P(l_0, l_1 \mid u, x)\P(t \mid x,u)\P(x,u)}{\int_u \P(l_0, l_1 \mid u, x)\P(t \mid x,u)\P(x,u) d u}  du; \label{eq:prob_cond_u}
\end{align}
where the indirectness of unmeasured confounding assumption is used in the second step. 

\bigskip

\section*{3. Bayesian Causal Factor Model}
\label{sec:model}
\bigskip \section*{\small Model Definition}
\label{sec:def_model}

To address the multivariate outcome, we propose an approach that adopt a factor regression model~\citep{carvalho2008high}, which uses factor analysis to model the potential outcomes given the treatment and the confounders.  
Rather than performing multiple univariate analyses, our factor analysis approach (i) improves statistical power by leveraging the correlation structure in the outcome, and (ii) facilitates the recovery of information from the unmeasured variable ${U}$, under the three cases considered in Figure~\ref{fig:dags}.

Specifically, we assume that for each unit $i\in \{1, \dots, n\}$, the outcome vector $\Y_i$, follows a treatment-specific probability density distribution $g_t(\cdot)$, with $t \in \{0,1\}$, where $g_t(\cdot)$ is modeled as a multivariate normal distribution 
\begin{gather}
    \{\Y_i \mid \x_i, t, \lt_{it} \} \sim g_t(\x_i, \lt_{it}), \notag \\
    g_t(\x_i, \lt_{it}) = \boldsymbol{\mu}_t + \B_t \x_i + \Lambda_t \lt_{it} + \boldsymbol{\xi}_{it}, 
    \quad \boldsymbol{\xi}_{it} \sim \Norm_q(0, \Psi_t). \label{eq:model}
\end{gather}
The outcome $\Y_i$ depends on treatment $t$, covariates $\x_i$, and treatment-specific latent factors $\lt_{it} \in \mathbb{R}^{J_t}$, where $J_t$ denotes the number of factors for treatment $t$ (assumed unknown). The matrix $\Lambda_t \in \mathbb{R}^{q \times J_t}$ represents treatment-specific factor loadings, \review{where the element $\lambda_{tjh}$ denotes the entry in row $j$ and column $h$}, and $\Psi_t \in \mathbb{R}^{q \times q}$ is assumed to be a diagonal matrix with elements $(\psi_{1t}, \dots, \psi_{qt})$ as in standard factor analysis literature \citep{thurstone1931multiple, anderson2004introduction}. 

Covariates $\X$ are  standardized such that $\x \sim \Norm_p(0,\rchi)$ with  diagonal elements of $\rchi$ equal to 1. The corresponding regression parameters $\B_t  \review{\in \mathbb{R}^{q \times p}}$ and intercept $\boldsymbol{\mu}_t$, for each treatment level $t$, have prior probability distributions, following the Bayesian paradigm. Among the reasonable prior distributions, we assume conjugate priors 
\begin{equation*}
    \boldsymbol{\mu}_t \sim \Norm_q (\boldsymbol{\mu}_{m_t},\Sigma_{m_t})\mbox{ and } \boldsymbol{\beta}_{tj} \sim \Norm_q (\boldsymbol{\mu}_{\beta_t},\Sigma_{\beta_t}),
\end{equation*}
where $\boldsymbol{\beta}_{tj}$ denotes the $j$-th row of the matrix $\B_t$ in \eqref{eq:model}, such that $\boldsymbol{\beta}_{tj} = \{\beta_{tj1}, \dots, \beta_{tjp} \}$ and $\beta_{tjk}$ is the element $(\B_t)_{j,k}$ with  \review{$j \in \{1, \dots, q\}$ and} $k \in \{1, \dots, p\}$. We assume that $\Sigma_{m_t}$ and $\Sigma_{\beta_t}$ are diagonal matrices for each $t$.


\bigskip \section*{\small The Role of the DDP in the Factor Scores}
\label{subsec: role_DDP}

In the literature, Bayesian statistical inference in factor analysis primarily focuses on priors for the factor loading matrix, whereas the prior for the factor score matrix is typically chosen as a standard Gaussian distribution. 

However, this prior is not suitable for handling the imputation of the missing data, as required in the potential outcomes framework for causal inference. Since accurate imputation of missing data is crucial, an informed choice of prior distribution is necessary.
To address this, in the following Section \ref{subsec:ddp}, we propose a prior distribution for the treatment-specific factor scores that takes advantage of the properties of the dependent Dirichlet process \citep[DDP;][]{quintana2020dependent}. 

The flexibility of the DDP has made it widely adopted in causal inference settings involving scalar outcomes.
For example, \citet{roy2024bayesian} leverage the enriched Dirichlet process to estimate causal effects for mediation analysis, while \citet{zorzetto2024confounder} specify a DDP prior for heterogeneous treatment effect.

In our factor model, our DDP prior has two characteristics: (i) it can written as infinite mixture of Gaussian distributions, and (ii) the probability of assignment to each mixture component depends on the covariates $X$---called cluster allocation in the DDP literature. 
Moreover, we specify a DDP prior that allows us to capture heterogeneity at both the factor and treatment levels. 
In fact, treatment-specific factor scores quantify the score/importance that each unit expresses on the corresponding factor under the specific treatment level. In the environmental application discussed in Section \ref{sec:application}, each factor score of a specific chemical indicates its contribution to explain the correlation among chemicals within the corresponding latent factor (reflecting underlying sources of metal and organic emissions). Notably, each factor score can vary depending on the presence or absence of wildfire smoke, reflecting how wildfire exposure alters the chemical composition of the air.

The characteristic (i) defines a flexible distribution for each factor score that captures the heterogeneity in the factor scores through the identification of different clusters, while characteristic (ii) associates the cluster allocation to specific values of the covariates.

This improves not only the estimation of factor scores compared with the standard Gaussian distribution as prior, but also their imputation under unobserved treatment levels. Specifically, for a given factor, the predictive posterior distribution of a factor score of a unit under the unobserved treatment level is similar to the posterior distributions of factor scores of units with similar covariate profiles, when observed under the same treatment. 
This data imputation process can be viewed as analogous to the matching process typically used in study design, where the criteria are driven by the covariates. However, our prior specification (Eq. \eqref{eq:prior_factors}) does not replace the study design in observational studies, as further discussed in the application section.


\bigskip \section*{\small Depend Mixture Distribution for Treatment-Specific Factor Scores}
\label{subsec:ddp}

We define the prior for each factor $l_{itj}$ as follows
\begin{gather}
      \{l_{it\review{h}} \mid \x_i,t \} \sim \int_\Omega {\mathcal H}(\cdot; \omega) dG_{\x_i}^{(t\review{h})}(\omega), \quad 
      G_{\x_i}^{(t\review{h})}\sim \Pi_{\x_i}^{(t\review{h})}, \label{eq:prior_factors}
\end{gather}
for $\review{h} \in \{1, \dots, J_t\}$ of units $i\in \{1,\dots. n\}$, for each $t\in \{0,1\}$.
The random probability measure $G_{\x_i}^{(t\review{h})}$ depends on confounders $\x_i$ and is specific to a treatment level $t$ and a factor $\review{h}$. Assuming its prior $\Pi_{\x_i}^{(t\review{h})}$ where $\Pi_{\x_i}^{(t\review{h})}$ is a DDP \citep{quintana2020dependent}, we define a dependent infinite mixture distribution as the probability density distribution of the factors. 

Consistently with standard factor analysis literature, we define the continuous density function ${\mathcal H}(\cdot;\omega)$, with parameter space $\Omega$, as a Gaussian kernel, where $\omega=(\eta, 1/\tau)$ are the location and scale parameters, respectively.

Following a single-atom DDP \citep{quintana2020dependent} characterization of the random measure $G_{\x_i}^{(t\review{h})}$, we can write:
\begin{equation}
G_{\x_i}^{(t\review{h})} = \sum_{r \geq 1} \pi_{r}^{(t\review{h})}(\x_i) \delta_{\mathbb{\omega}^{(t\review{h})}_{r}},
\label{eq:asdiscreteG}
\end{equation}
for $t=\{0,1\}$ and $\review{h}=\{1, \dots, J_t\}$, where $\{\pi_{r}^{(t\review{h})}(\x_i)\}_{r\geq 1}$ and $\{\omega^{(t\review{h})}_{r}\}_{l\geq 1}$ represent infinite sequences of random weights and random kernels' parameters, respectively. Notably, both random sequences depend on the level of treatment, $t$, and factor, $\review{h}$, while the weights also depend on the values $\x_i$ of the confounders. 

The elements of the sequence of random parameters $\{\omega_{r}^{(t\review{h})}\}_{r\geq 1}$ are independent and identically distributed, where the $r$-th element of the sequence is the joint parameters of the $r$-th Gaussian distribution in the mixture model in \eqref{eq:prior_factors}, so that $\omega_r^{(t\review{h})}=(\eta_r^{(t\review{h})}, 1/\tau_r^{(t\review{h})})$.
Following the stick-breaking representation \citep{sethuraman1994constructive}, the sequence of dependent weights is defined as
\begin{equation*}
    \pi_{r}^{(t\review{h})}(\x_i)  = V^{(t\review{h})}_{r}(\x_i)\prod_{g<r}\{1-V_{g}^{(t\review{h})}(\x_i)\},
\end{equation*}
where $\{V_{r}^{(t\review{h})}(\x_i)\}_{r\geq 1}$ are $\{0, 1\}$-valued independent stochastic processes.

Among the dependent nonparametric processes, the Dependent Probit Stick-Breaking (DPSB) \citep{rodriguez2011nonparametric} defines the process $\{V_{r}^{(t\review{h})}(\x_i)\}_{r\geq 1}$ as follows:  
\begin{gather}
    V_{r}^{(t\review{h})}(\x_i)  = \Phi(a_{r}^{(t\review{h})}(\x_i)), \quad 
    a_{r}^{(t\review{h})}(\x_i) \sim \Norm( \alpha_{0r}^{(t\review{h})} + \alpha_{1r}^{(t\review{h})}\x_{1i}+\dots + \alpha_{pr}^{(t\review{h})}\x_{pi},1),
    \label{eq:dpsb}
\end{gather}
where $\Phi(\cdot)$ is the Probit function and $\{a_{r}^{(t\review{h})}(\x_i) \}_{r\geq 1}$  have Gaussian distributions whose mean is a linear combination of the $p$ covariates $\x_i$.

This prior distribution for treatment-specific factors \eqref{eq:prior_factors} allows us to introduce a latent categorical variable $S_{it\review{h}}$, for each factor $\review{h} \in \{1, \dots, J_t\}$ treatment $t\in \{0,1\}$ unit $i\in \{1,\dots. n\}$, describing clusters of units defined by heterogeneous responses to the treatment level $t$. 
Assuming $\Pr(S_{it\review{h}}= r) = \pi_r^{(t\review{h})}(\x_i)$, then the prior distribution of the treatment-specific factors \eqref{eq:prior_factors} can be rewritten, conditional on $S_{it\review{h}}$, as
\begin{equation}
   \{ \lt_{it\review{h}} \mid S_{it\review{h}}=r, \boldsymbol{\omega} \} \sim \Norm(\eta_r^{(t\review{h})}, 1/\tau_r^{(t\review{h})}).
   \label{eq:lat_cond_weitgh}
\end{equation}
For each treatment $t$ and cluster $r$, we assume $\tau_r{th}$ fixed to 1 and the prior distribution for the location parameter as $\eta_r^{(th)} \sim \Norm(0,1)$.


The proposed prior for treatment-specific factor scores offers several strengths, as outlined in Section \ref{subsec: role_DDP}. Additionally, it serves as a powerful tool for inferring information about the relationships among the multiple outcomes, as governed by unobserved features $U$. \review{A detailed discussion of these statistical guarantees is provided in the Supplementary Material.}

\bigskip \section*{\small Shrinkage Priors for Factor Loading}
For the factor loading matrix, we adopt the well-known shrinkage prior introduced by \citet{bhattacharya2011sparse}. 
For each treatment-specific factor loading $\lambda_{t j h}$, where treatment level $t\in\{0,1\}$, the factor $h=1, \ldots, \infty$ and the outcome variable $j\in \{1 \ldots p\}$:
\begin{gather*}
    \lambda_{t j h} \mid k_{t j h}, \iota_{t h} \sim N\left(0, \theta_{t j h}^{-1} \iota_{t h}^{-1}\right), \; 
    \theta_{t j h} \sim \operatorname{Gamma}\left(\nu_{t} / 2, \nu_{t} / 2\right),  \\
    \iota_{t h} =\prod_{l=1}^{h} \delta_{t l},  \; 
    \delta_{t 1} \sim \operatorname{Gamma}\left(a_{1_t},1\right), \; \delta_{t l} \sim \operatorname{Gamma}\left(a_{2_t},1\right) \forall l \geq 2. 
\end{gather*}

We follow the default hyperparameter settings proposed by \citet{bhattacharya2011sparse}.

\bigskip

\section*{4. Simulation Study}
\label{sec:simulations}
We evaluated the performance of the proposed model through an extensive simulation study. The primary objective is to assess its ability to estimate causal effects, with a particular focus on controlling the variability of {\bf SATE} and adjusting the bias induced by the unmeasured variable $U$. We compare our causal factor model to state-of-the-art flexible models in causal inference---causal BART by \cite{hill2011bayesian} and Bayesian causal forest (BCF) by \cite{hahn2020bayesian}---as well as a standard factor model employing a standard normal prior for factor scores and adapted the causal inference setting. Specifically, the standard factor model is defined as in \eqref{eq:model}, but with priors $ \lt_{it} \sim \Norm(0,1)$ for each unit $i$ and treatment level $t \in \{0,1\}$. This setup allows us to compare and evaluate the impact of the proposed DDP prior on treatment-specific factor scores.
While BART and BCF are estimated separately for each element of the outcome variable, both the standard factor model adapted to causal inference and our proposed approach explicitly account for the multivariate nature of the outcome.

We consider four simulation scenarios. The first three explore different relationships between the unmeasured variable $U$ and the confounders $\X$, corresponding to the graphical representations in Figure~\ref{fig:dags}. 
The fourth scenario is designed to closely mimic the real dataset used in our application in Section \ref{sec:application}. Details of simulation-generating processes are provided in Table~\ref{table:simulation_characteristics} \review{and fully described in the Supplementary Materials. }

\begin{table}[h!]
\caption{Parameters and distributions for the simulation scenarios.}
\label{table:simulation_characteristics}
\centering
\scriptsize
\scalebox{0.9}{
\begin{tabular}{ccccc}
    & Scenario 1 & Scenario 2 &Scenario 3 & Scenario 4 \\
\hline
    \multicolumn{4}{l}{{\bf Dimensions:}} \\
$n$ & 500 & 500 &500 & 3426\\
$J_t$& $(3,3)$ & $(3,3)$ & $(3,3)$ & $(3,3)$ \\
$q$& 10 & 10 & 10 & 27 \\
$p$& 4 & 4 & 4 & 28 \\
\hline
\multicolumn{4}{l}{{\bf Variables:}} \\
$U$ & $U \sim N(0,2)$ & $U \sim N(f_U(X_{1:4}),0.5)$ &$U \sim N(0,2)$& - \\
$\X$ & $X_{1:2} \sim N(0,1)$ & $X_{1:2} \sim N(0,1)$ & $X_{k} \sim N(f_k(U),1),\, k=1,2$ & observed $\X$\\
     & $X_{k} \sim Be(\pi_k),\, k=3,4$& $X_{k} \sim Be(\pi_k),\, k=3,4$& $X_{k} \sim Be(\pi_k),\, k=3,4$& \\
$T$ &  $T \sim Be(f_{T}(X_{1:4},U))$ & $T \sim Be(f_{T}(X_{1:4},U))$ & $T \sim Be(f_{T}(X_{1:4},U))$ & observed $T$ \\
\hline
\multicolumn{4}{l}{{\bf Factors:}} \\
Clusters & \multicolumn{3}{c}{3 or 2 cluster for each treatment-specific factors: $C_{ith}  := f_C(\X_{1:4})$} & $C_{ith}  := f_C(\X_{1:27})$ \\
$l_{it}$ &\multicolumn{3}{c}{$\{l_{ith} \mid C_{ith}=c\} \sim N(\mu_{tc}+\gamma_{tc}U, 1), \; h=\{1, \dots, p\}$}& $\{l_{itj} \mid C_{ith}=c\} \sim N(\mu_{tc}, 1)$\\
$\Lambda_t$&\multicolumn{3}{c}{matrices with $25\%$ of zeros and nonzeros elements generated}& estimated $\Lambda_t$\\
  &\multicolumn{3}{c}{by $\lambda_{th}\sim \pi Unif[-1,-0.8]+(1-\pi) Unif[0.8,1]$, with $\pi\sim Be(0.5)$}& in observed data\\
\hline
\multicolumn{4}{l}{{\bf Outcome model:}} \\
 $B_t$ & \multicolumn{4}{c}{ $\beta_{ht} \sim Unif[-3 +t, 2+t], \; t=0,1, \; h= 1, \dots, p$} \\
$\epsilon_t$ & \multicolumn{4}{c}{ $\xi_{it} \sim Unif_p[0,1], \; t=0,1$} \\
${\bf Y}$ & \multicolumn{4}{c}{ $Y_i(t) = B_tX +\Lambda_t l_{it} + \xi_{it}, \; t=0,1 \; i=1, \dots, n$} \\
&&&& \\
&&&& \\
\end{tabular}}
\end{table}

For each of the four scenarios, we generated \review{100} collections of datasets. Causal effects are estimated using the \texttt{R} package \texttt{bartCause} to estimate with the BART model, the \texttt{bcf} code available on GitHub for BCF estimation, and our own Gibbs sampler for the factor models, available on GitHub at \href{https://github.com/dafzorzetto/BayesCausalFactor}{dafzorzetto/BayesCausalFactor}. The same codebase includes the implementation of the standard factor model.

\begin{figure}[h!]
\centerline{
    \includegraphics[trim={0cm 14cm 0cm 0cm},clip,width=190mm]{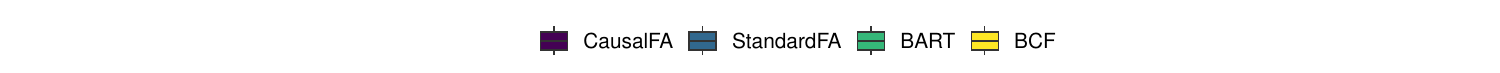}}
\centerline{
    \includegraphics[trim={0cm 0cm 0cm 9.5mm},clip,width=70mm]{plots/2nd_submission_sim/Scenario_1_Bias.pdf}
    \includegraphics[trim={0cm 0cm 0cm 9.5mm},clip,width=70mm]{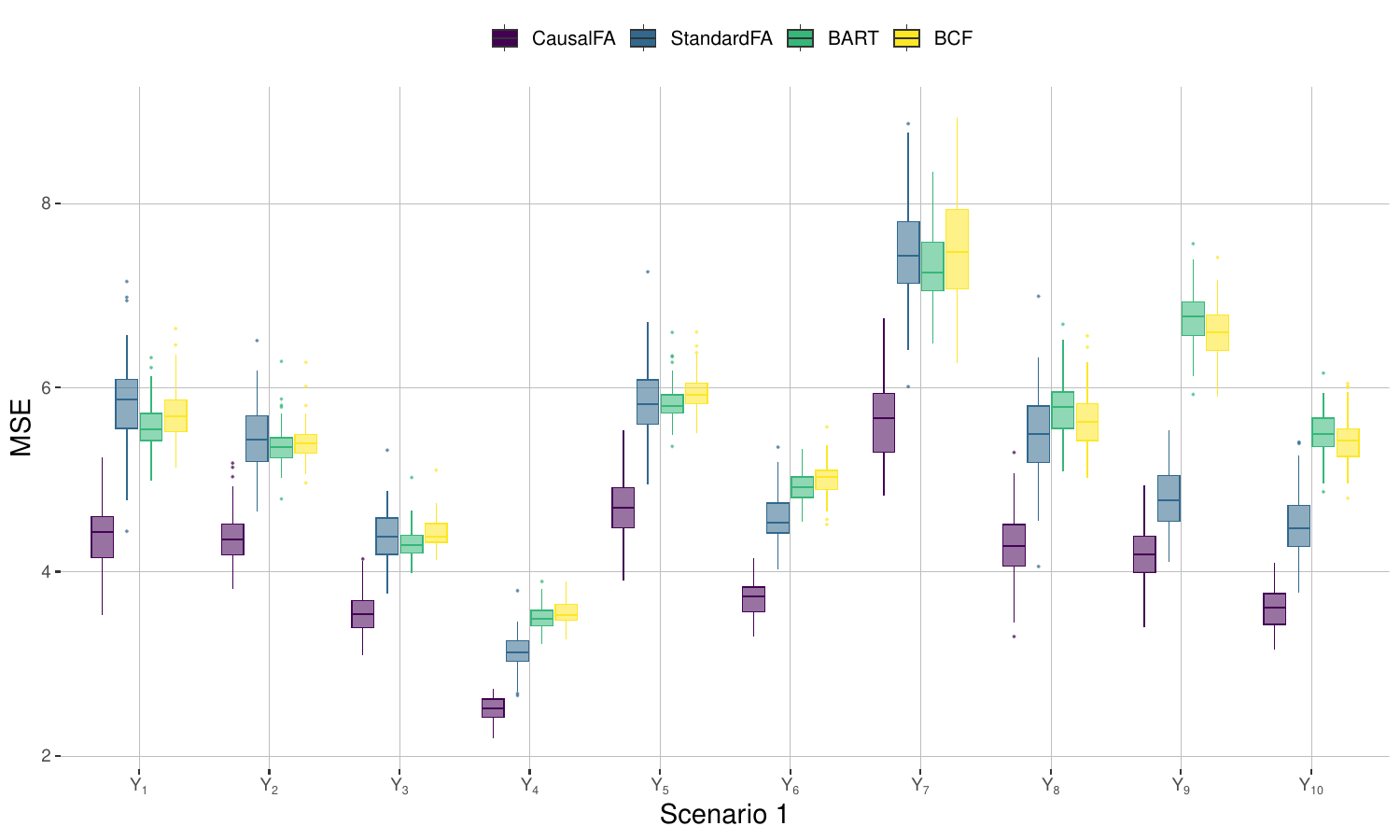}}
\centerline{
    \includegraphics[trim={0cm 0cm 0cm 9.5mm},clip,width=70mm]{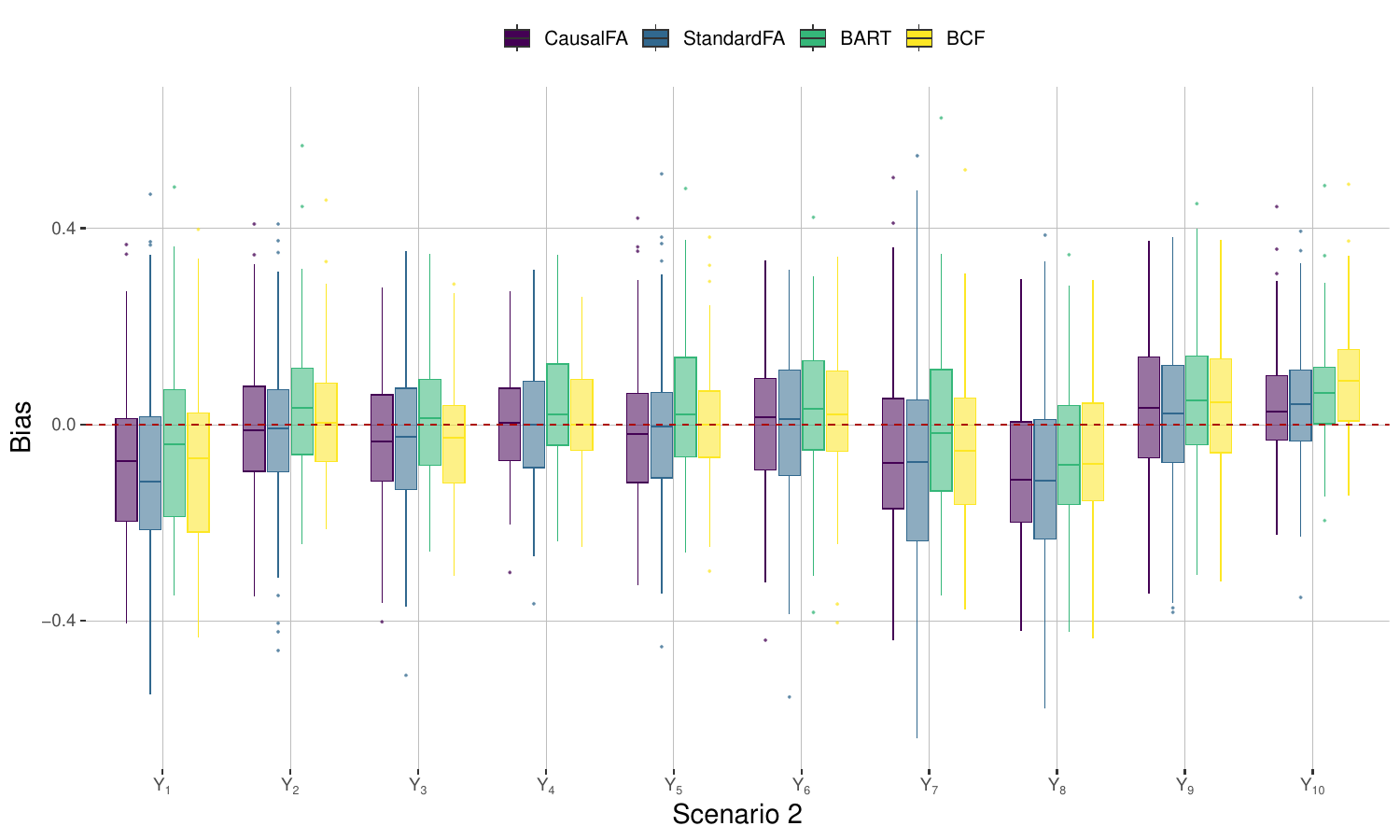}
    \includegraphics[trim={0cm 0cm 0cm 9.5mm},clip,width=70mm]{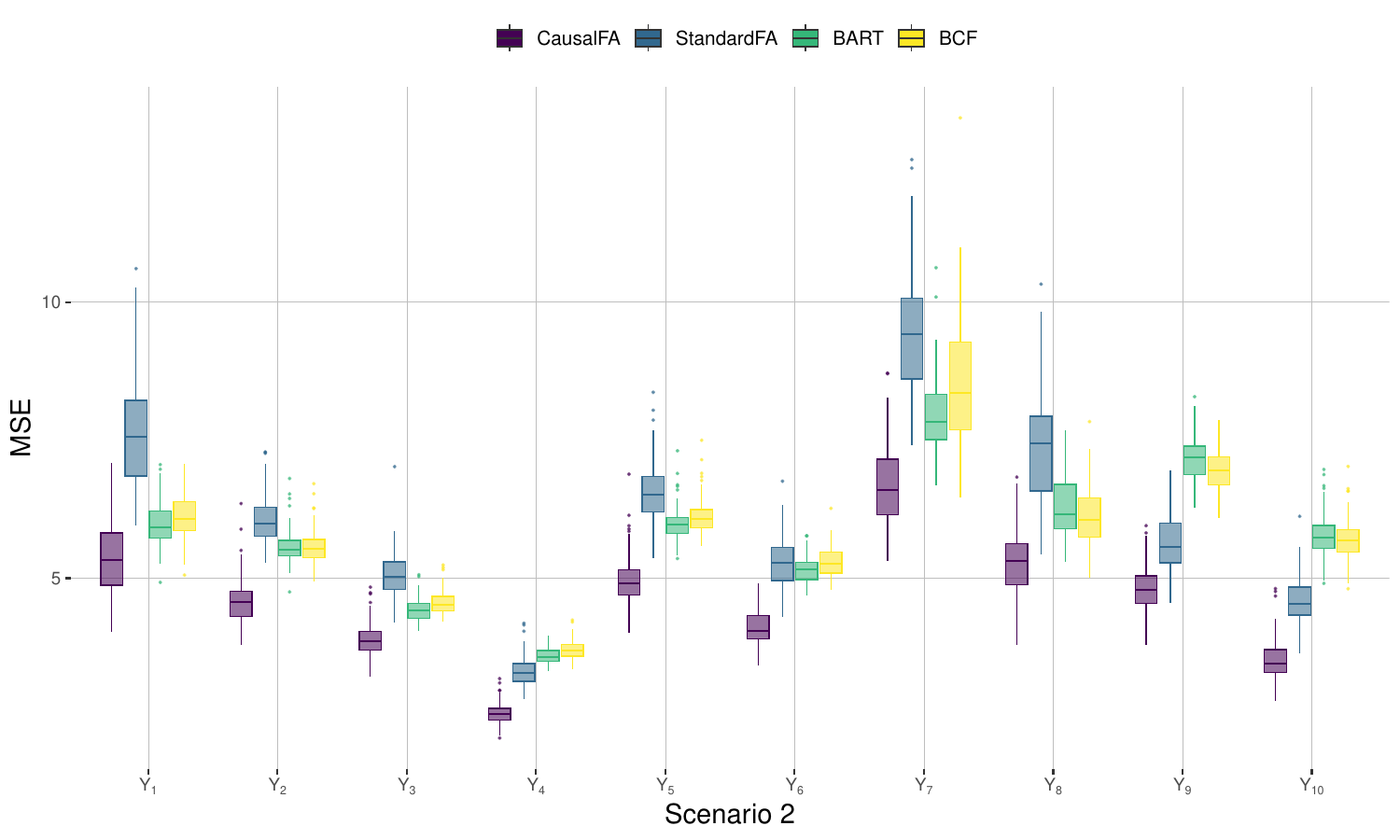}}
\centerline{
    \includegraphics[trim={0cm 0cm 0cm 9.5mm},clip,width=70mm]{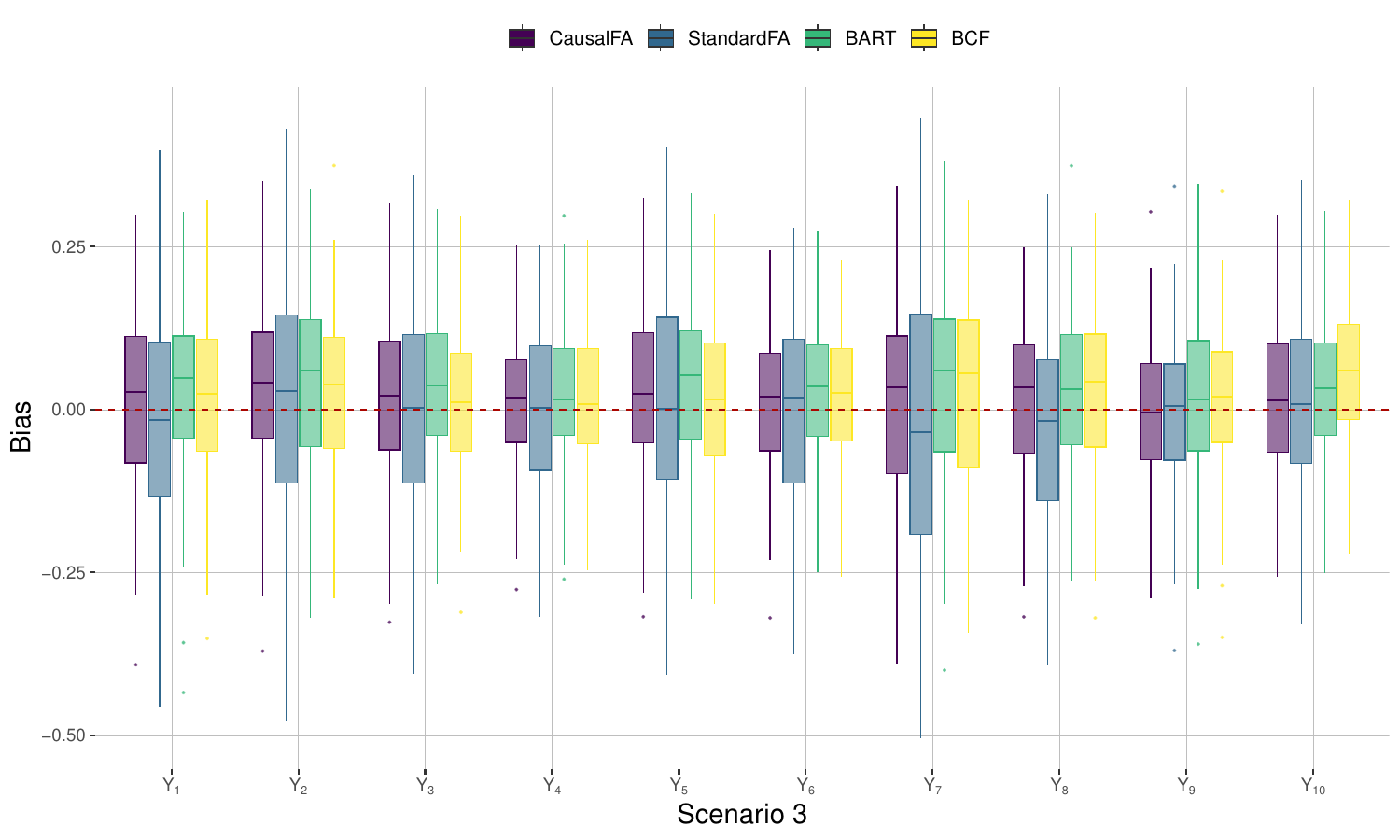}
    \includegraphics[trim={0cm 0cm 0cm 9.5mm},clip,width=70mm]{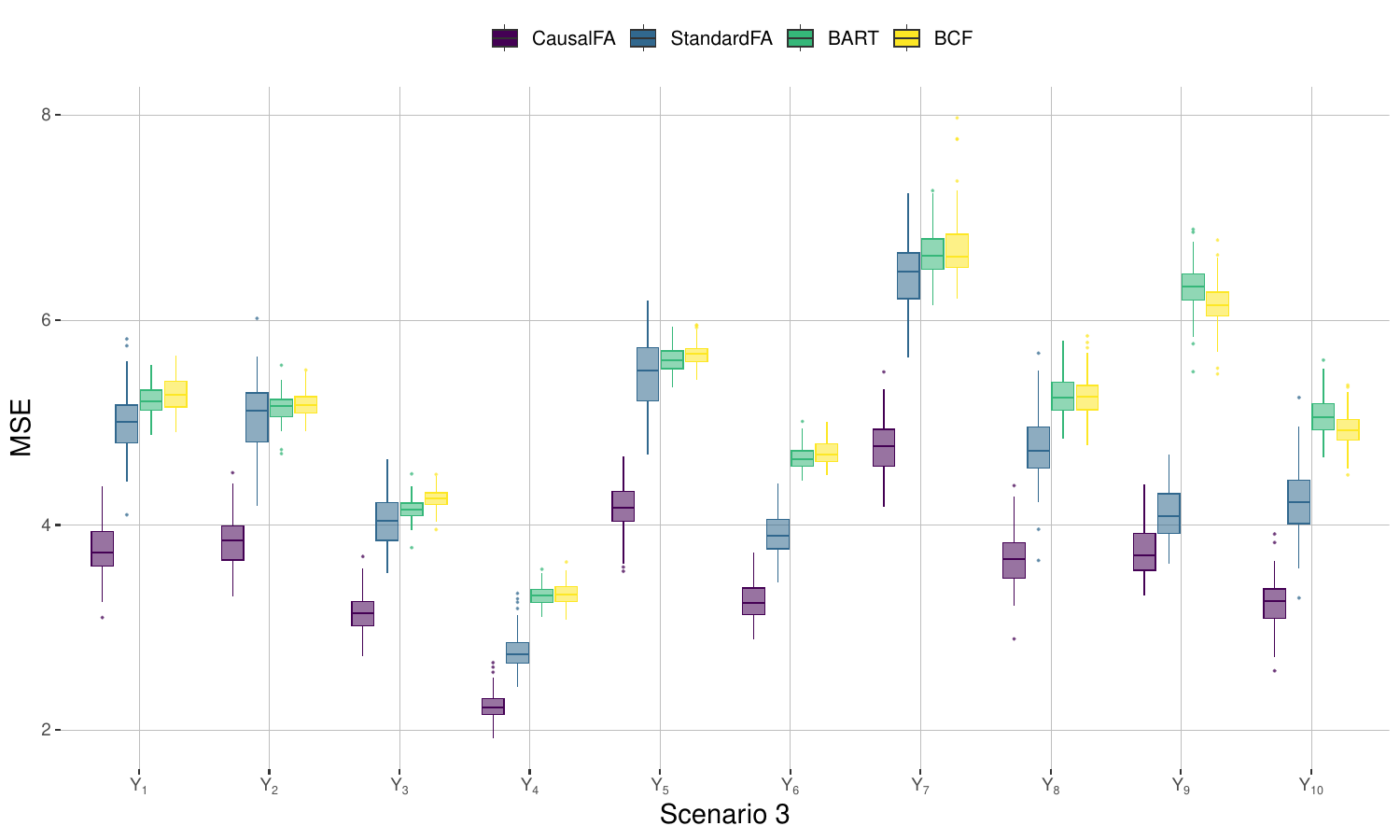}}
\centerline{
    \includegraphics[trim={0cm 0cm 0cm 9.5mm},clip,width=70mm]{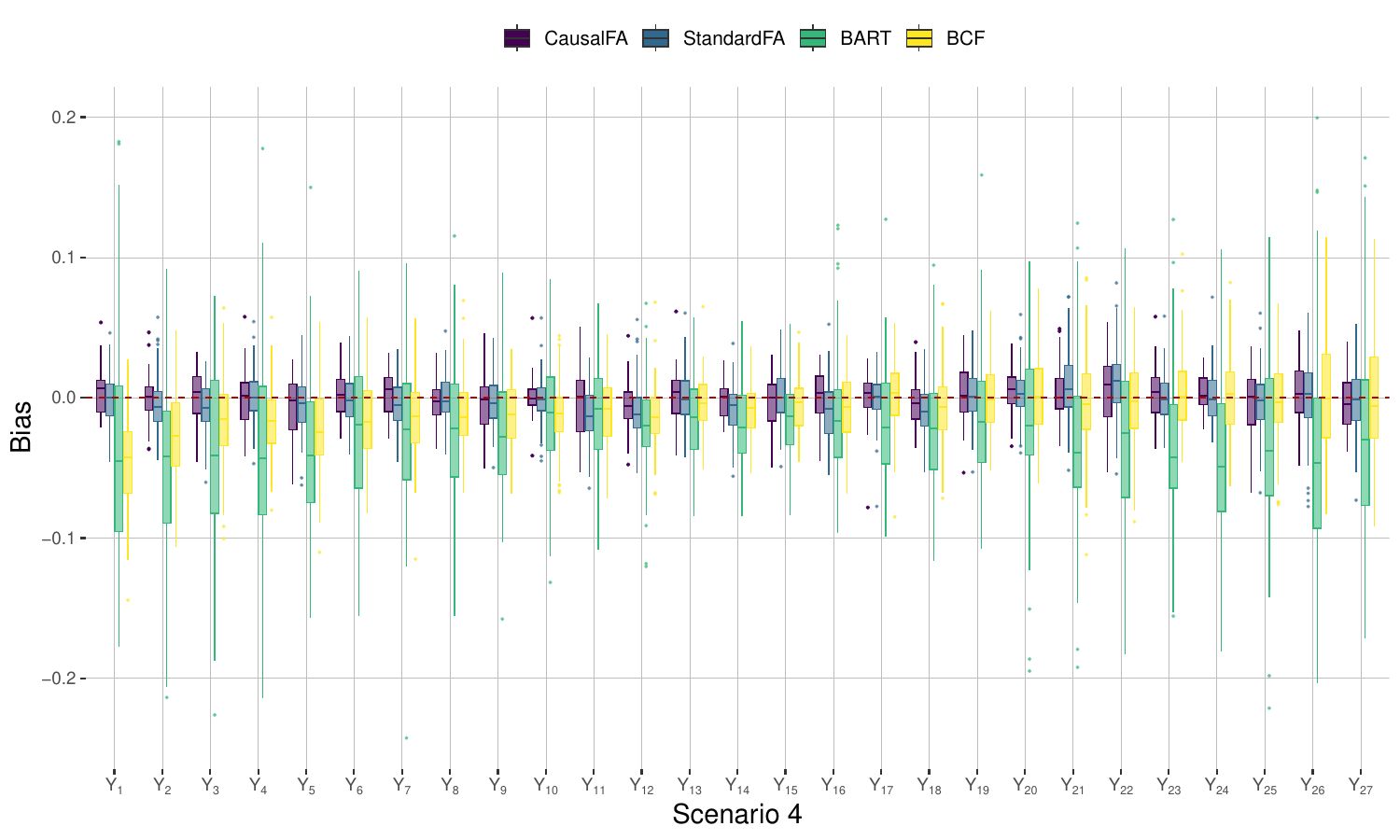}
    \includegraphics[trim={0cm 0cm 0cm 9.5mm},clip,width=70mm]{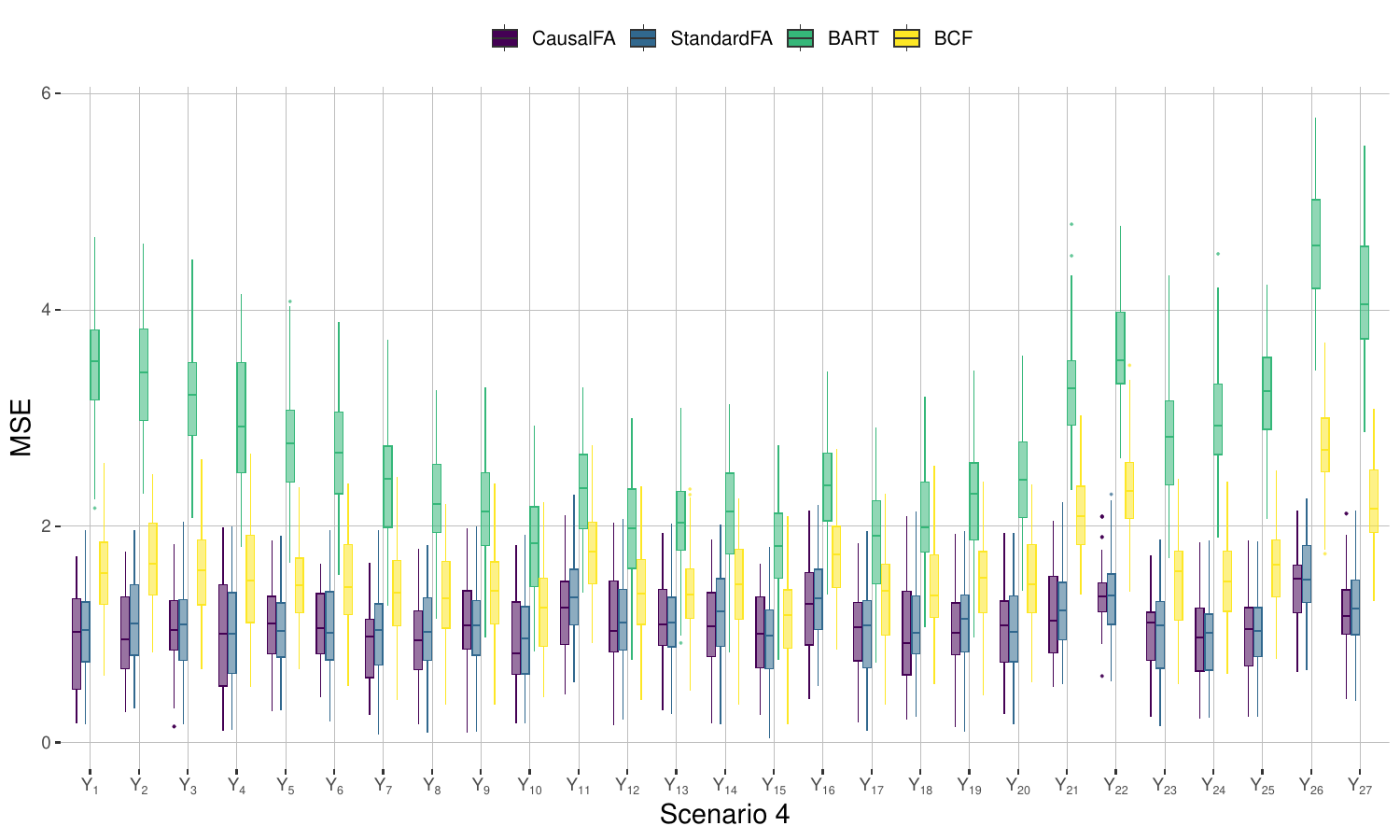}}
\caption{\review{Bias (left) and mean square error (MSE, right) across simulated scenarios. Results are shown for our proposed model, standard factor model, causal BART, and BCF. \label{fig:sim}}}
\end{figure}

We evaluate the performance of the models according to the bias mean and square error (MSE) of \kATE\, estimation for each $k =\{1, \dots, 10\}$ in Scenarios 1–3, and $k=\{1,\dots,27\}$ in Scenario 4. The results are summarized in Figure~\ref{fig:sim}. 
Our proposed model (\textit{CausalFA}) outperforms the competitors in both bias and MSE, showing unbiased estimations of causal effects and a smaller variability.  Specifically, BART and BCF exhibit higher values for the MSE, with particular ample boxplots for the BART. The factor model with standard normal prior (\textit{StandardFA}) estimates causal effects with bias, highlighting a critical weakness of standard prior for factor scores and the benefit of the DDP prior.
Only our model provides unbiased causal effect estimates across all simulated scenarios, effectively adjusting for unmeasured variables via latent factors and their priors. \review{Moreover, Figure C.2 in the Supplementary Materials demonstrates good coverage properties for our proposed model. }

\review{Additional simulation setting are analyzed in the Supplementary Materials, Appendix C.}

\bigskip

\section*{5. Environmental Application}
\label{sec:application}

Wildfires have become a major environmental and public health concern, significantly deteriorating air quality through the release of smoke and associated pollutants. Although previous studies have examined the broad effects of wildfire smoke on particulate matter (\PMns), a detailed understanding of its causal effect on specific chemical species in the environment remains limited. The recent work of \citet{krasovich2025influence} laid the foundation for studying wildfire smoke and chemical variation in the United States, highlighting strong interconnections between them. However, their analysis is based on temporal correlations adopting a linear modeling approach for each individual chemical. Our study addresses two distinct objectives employing a novel methodology. First, we aim to answer the causal question: \textit{What is the causal effect of wildfire smoke on each of the chemicals on \PMns?} Second, we explicitly account for the correlation between chemicals, providing robust estimations through our proposed causal factor model.


\bigskip \section*{\small Data}
The data analyzed are merged from different sources, following a strategy similar to \citet{krasovich2025influence}. We consider the $280$ air pollution monitors across the United States, operated by the National Park Service and the U.S. Forest Service. The data are collected from the Environmental Protection Agency's (EPA) \PM Chemical Speciation Network (CSN) and the Interagency Monitoring of Protected Visual Environments (IMPROVE) program, and stored in the Federal Land Manager Environmental Database \citep{FED_improve}. We select the same $27$ chemical species considered in \citet{krasovich2025influence}, divided in alkaline-earth metals, alkali metals,
transition metals, metalloids, other metals, nonmetals, halogens, and Organics. The full list of chemical species is provided in the Supplementary Materials.

Each monitor records daily information on the level of each chemical species. The dataset includes data for one out of every three days. The IMPROVE and CSN monitoring data were preprocessed to replace data flagged as unacceptable quality with interpolated values or, when appropriate, a value equal to 1/2 of the minimum detection limit (less than 3\% of values were interpolated). To adjust for methodological differences between the CSN and IMPROVE monitors, we include the variable \textit{monitor\_type} as a confounder.

Wildfire smoke \PM concentration are derived from the daily prediction of \citet{childs2022daily}, available from the GitHub repository \href{https://github.com/echolab-stanford/daily-10km-smokePM}{echolab-stanford/daily-10km-smokePM}. 
\review{We define the binary treatment variable ``smoke day'' as follows: for each monitor-day unit, treatment level $t = 1$ (a ``smoke day'') indicates that non-zero smoke-related \PM is predicted at that monitor location for that day. Conversely $t = 0$ (a ``non-smoke day'') indicates that no smoke-related \PM is predicted. Further details are in the Supplementary Materials.}

\review{In our analysis, SUTVA assumes that the wildfire smoke \PM concentration at a given location is only dictated by the smoke at that location and not by smoke at other locations. Essentially, once we know the wildfire smoke at a given location, whichever wildfire(s) it might have originated from or the smoke at neighboring locations is not relevant to the chemicals at that location.}

As confounders, we collect weather information (from the Harvard Dataverse \citep{harvarddataverse}) for U.S. ZIP Code Tabulation Areas (ZCTA) at daily resolution and census data (via the \texttt{tidycensus} \texttt{R} package \citep{walker2021package}). Weather variables include daily values for near-surface air temperature, maximum near-surface air temperature, precipitation, minimum and maximum near-surface relative humidity, surface downwelling solar radiation, and wind speed and direction at $10m$. Census variables include population, percentage of males, percentage of different ethnicities (White, Black and African American, Asian, and other), median household income, total number of housing units, and poverty status. Both weather information and census data are matched with the monitor locations.

Our analysis focuses on July–September 2014, when wildfire smoke is most prevalent in the United States \citep{krasovich2025influence}. The observational unit is a monitor-day, for a total of $7,467$ units. The study design, described in detail in the Supplementary Materials, employs a 1-to-1 nearest neighbor propensity score matching, yielding a balanced sample of $3,426$ units, and improving the balance of the covariates. The causal effects, estimated in the following section, target the sample obtained through the matching procedure.

\bigskip \section*{\small Results}
\begin{figure}[h!]
\centerline{
    \includegraphics[width=0.9\textwidth]{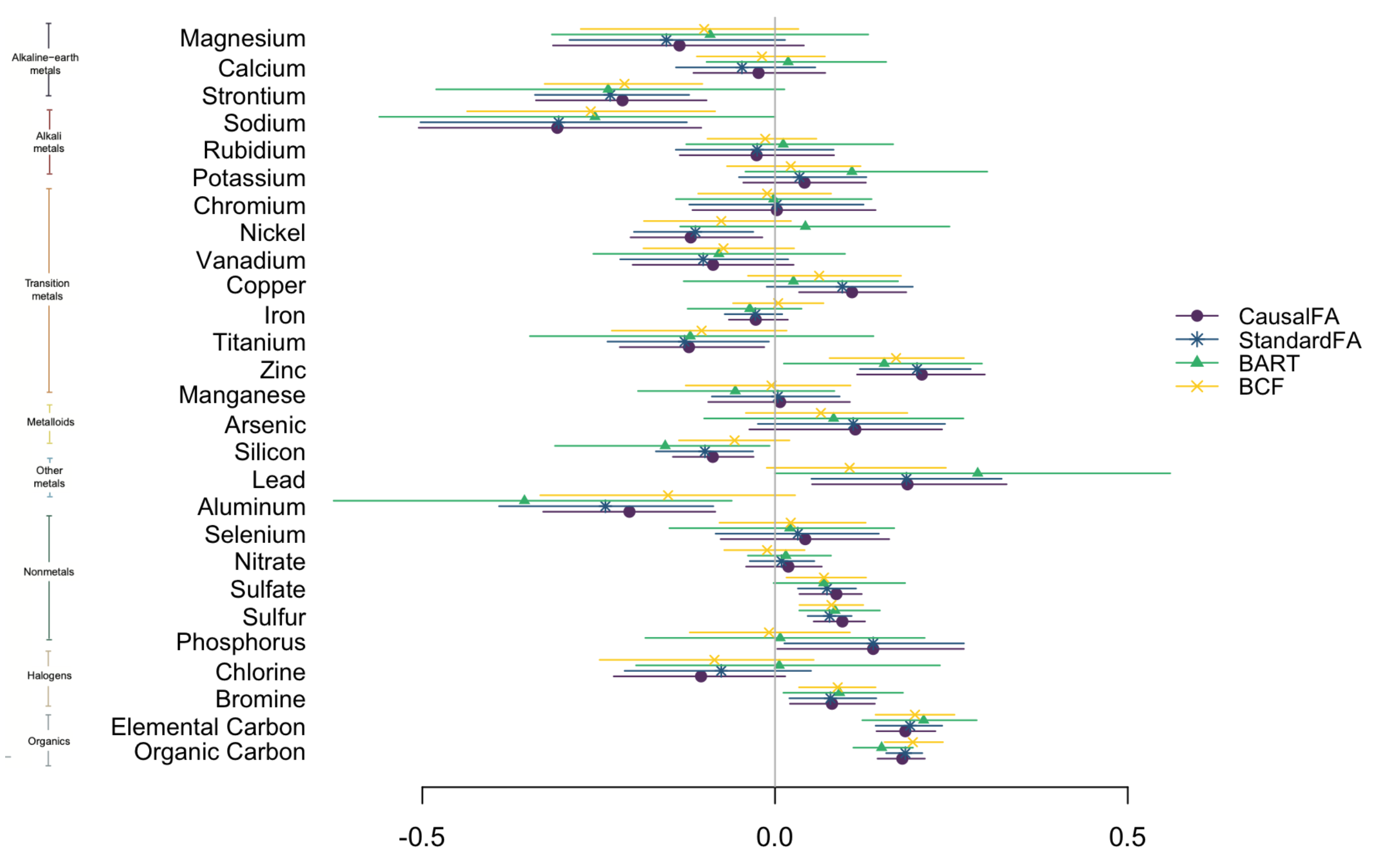}
}
\caption{Estimated causal effects from four models: our proposed causal factor model, factor model with standard Gaussian prior for factor scores, BART, and BCF. Dots show the median of causal effects; lines show the corresponding $90\%$ credible interval. \label{fig:app_CE}}
\end{figure}

Similarly to the simulation study, we estimate the causal effect of wildfire smoke on $27$ chemical species using four models: our proposed causal factor model, the factor model with a standard Gaussian distribution for the factor scores, BART, and BCF. The results are compared in Figure~\ref{fig:app_CE}.

All four models consistently identify a positive causal effect of wildfire smoke on zinc, bromine, and the two organic elements---elemental and organic carbon---indicating that wildfire smoke increases the levels of these chemicals. This finding aligns with previous studies: \citet{odigie2014trace} observed increased zinc concentrations following wildfire smoke in California, \citet{liu2014wildland} reported elevated carbon levels and their negative impact on ecosystems, and \citet{li2023large} found higher bromine concentrations during wildfires in urban areas.
However, the models do not fully agree on the detection of significant negative causal effects, i.e. a reduction in chemical concentrations due to wildfire smoke, particularly for strontium, sodium, nickel, titanium, silicon and aluminum. This discrepancy arises from the wide credible intervals produced by BART, which often include zero, and, in some cases, by BCF as well, limiting detection of significant effects. 
As shown in the simulation study, these two models tend to estimate wide variability, particularly in highly correlated outcomes. Similarly, for positive effects, only our proposed model has credible intervals that do not include zeros for cooper, lead, sulfate, and phosphorum. This is consistent with previous studies, as \citet{odigie2014trace} observed notable increases in these metals, while \citet{spencer1991phosphorus} reported elevated phosphorus concentrations following wildfires.

\begin{figure}[h!]
\centerline{
    \includegraphics[width=1\textwidth]{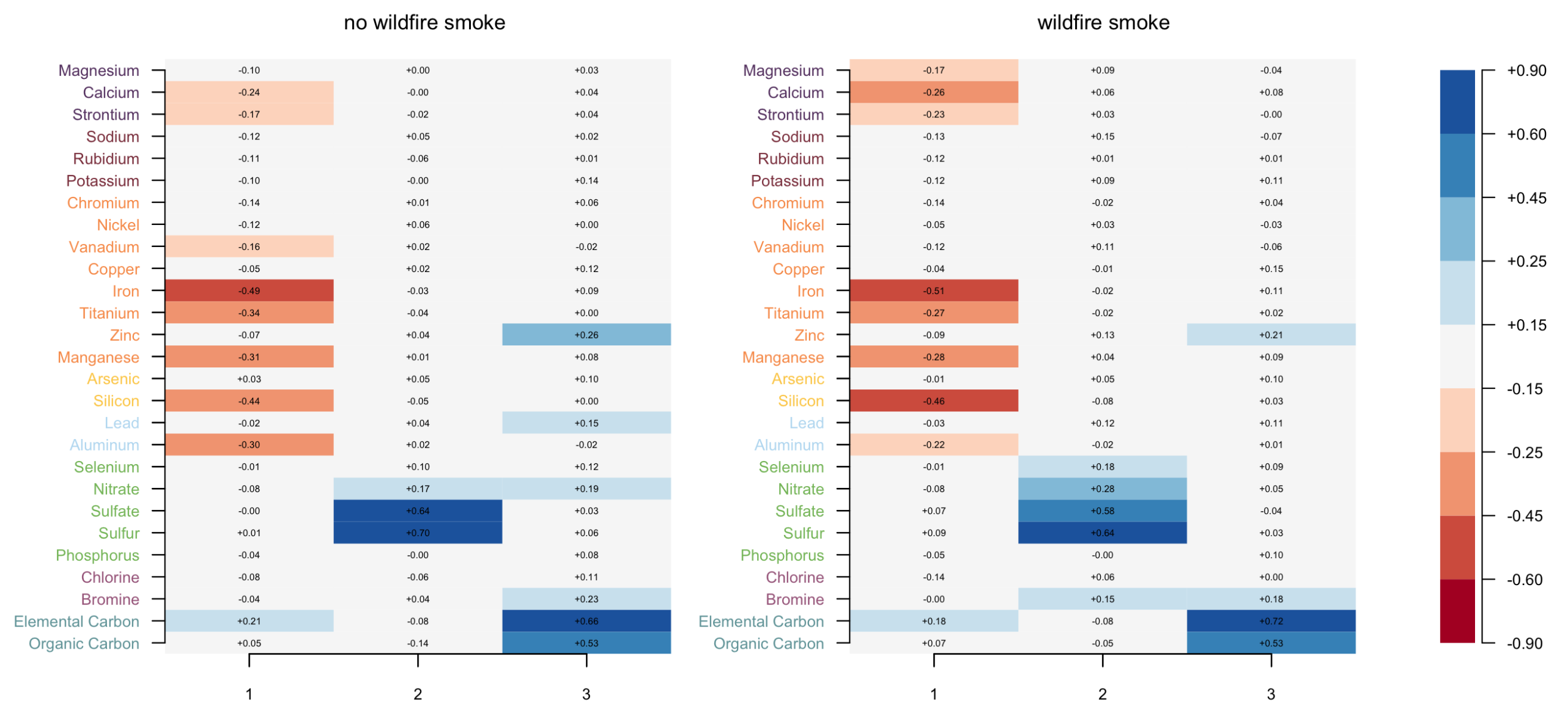}
}
\caption{Treatment-specific factor loadings: (left) without wildfire smoke and (right) with wildfire smoke. Colors of the chemical species names indicate chemical groupings. \label{fig:app_factors}}
\end{figure}

The different distributions of the causal effects identified by the four models can be better understood by examining the factor loading matrix, reflecting the correlations among the 27 chemical species. Our Bayesian causal factor model framework enables us to analyze these relationships and assess how wildfire smoke alters them by estimating treatment-specific factors---i.e., distinct factors for the presence and absence of wildfire smoke. After applying a varimax transformation \citep{kaiser1960application}, we identify $3$ factors for each treatment level, explaining a total of $85\%$ and $88\%$ of the variance, respectively for wildfire smoke and non-smoke days (Figure \ref{fig:app_factors}).
In both treatment levels, the first factor captures strong correlations among transition metals, metalloids, and other metals. 
Notably, titanium, silicon, and aluminum, chemicals whose concentrations drive the Factor 1, are detected to have significant wildfire smoke effects by only the factor models (\textit{CausalFA} and \textit{StandardFA}), highlighting their ability to include correlation across the outcome components and narrow uncertainty and more robust conclusions.
Factor 2 captures strong correlations among non-metals, while Factor 3 reflects associations among organic elements. In both cases, wildfire smoke is estimated to increase factor levels. This may be plausibly related to wildfires burning natural materials such as forests, releasing elemental and organic carbon into the air \cite{liu2014wildland}. Likewise, non-metals, commonly found in soil, exhibit strong positive correlations with wildfire events \citep{spencer1991phosphorus}.
Furthermore, the presence of wildfire smoke appears to enhance correlations among alkaline-earth metals, transition metals, and metalloids within Factor 1, particularly among chemicals that share the same negative causal effect. Similarly, in Factor 2, wildfire smoke strengthens correlations among non-metals, while in Factor 3, it disrupts the correlation between organic chemicals and other elements, reinforcing their internal associations.
These structural shifts underscore the utility of our model in uncovering the latent structure of treatment.


\bigskip

\section*{6. Discussion}
\label{s:discuss}

This work introduces two key innovations. First, from a factor analysis perspective, we propose an infinite mixture distribution for the factor scores, addressing a critical limitation in the existing literature that has historically focused on factor loadings while largely overlooking the role of latent factor scores. Second, within the causal inference framework, we develop a multivariate outcome setup that enables the sharing of latent structure across outcome components, allowing adjustment for unmeasured variables potentially affecting components of the latent structure. We showed how this can improve the estimation of vector-valued causal effects.

By combining these two contributions, the proposed Bayesian causal regression factor model demonstrates superior performance in causal effect estimation compared to state-of-the-art methods in both the factor analysis and causal inference domains, also correcting the bias in the causal effect estimates. Specifically, it outperforms standard factor models that rely on Gaussian priors for factor scores, which fail to characterize heterogeneity in factor scores and consequently impute the missing outcome.  At the same time, it outperforms established methods in causal inference such as BART and BCF, which, despite their flexibility, lack the ability to share latent structure across correlated outcome components. Moreover, BART and BCF exhibit higher MSE values, likely due to their independent estimation of each causal effect.
However, future extensions could enhance our causal factor model by leveraging the flexibility of BART priors \cite{chipman2010bart}. Additionally, alternative priors can be considered for both the covariate regression---such as mixtures---and the factor loadings, including the 
generalized sparse prior of \citet{schiavon2022generalized}.

While the proposed approach can be broadly applied to various real-world datasets for studying the effects of treatments or exposures on outcomes composed of correlated elements, our application in this paper focuses on a critical environmental issue. Specifically, we aim to estimate the causal effect of wildfire smoke on the chemical composition of \PMns, motivated by the significant role that fine particulate exposure plays in human health. Our findings not only identify the chemicals that are positively or negatively affected by wildfire smoke but also reveal significant effects that other models fail to detect, due to its ability to borrow strength across outcomes. Moreover, our method effectively identifies the latent factors that drive the behavior of chemical components across different types of metals and non-metals, capturing treatment-specific patterns across classes of metals, nonmetals, and organics. This capability enriches our understanding of how wildfire smoke alters the composition of air quality in complex and systematic ways.
Further research could extend this work by modeling multivariate outcomes under exposure to wildfire smoke over time, with methodologies similar to \citet{samartsidis2024bayesian} or by accounting for the positive continuous nature of the prediction of \PM induced by wildfire smoke, as discussed in \citet{krasovich2025influence}, though in a non-causal framework.

We hope that this study highlights the importance of factor scores---not only in causal inference, as recently shown by \citet{zorzetto2024sparse} and \citet{bortolato2024adaptive}---and indicates the potential of factor models to quantify treatment effects in causal inference, particularly in settings where outcomes are high-dimensional and strongly correlated.


\bigskip

\section*{\small Acknowledgements}

The authors thank Ting Zhang and Lucas Henneman for providing the preprocessed chemical speciation data from CSN and IMPROVE monitors. RDV was supported by the US National Institutes of Health, NIGMS/NIH COBRE CBHD P20GM109035. 

\bigskip

\bibliographystyle{chicago} 
\bibliography{biomsample}

\makeatletter
\newenvironment{stdfigure}
  {\@float{figure}}
  {\end@float}
\makeatother

\newpage
\appendix
\counterwithin{figure}{section}
\counterwithin{table}{section}
\setcounter{equation}{0} 

 \pagenumbering{arabic}
    \setcounter{page}{1}

{\centering \bf \large SUPPLEMENTARY MATERIALS \\
`Multivariate Causal Effects: a Bayesian Causal Regression Factor Model'\\}
\vspace{0.6cm}

{\centering Dafne Zorzetto$^{1,*}$, 
Jenna Landy$^{2}$, Corwin Zigler$^{3}$, Giovanni Parmigiani$^{2,4}$, and \\
Roberta De Vito$^{1,3}$\\
\vspace{0.4cm}

$^{1}$ Data Science Institute, Brown University, Providence, Rhode Island, U.S.A. \\
$^{2}$ Department of Biostatistics, Harvard University, Cambridge, Massachusetts, U.S.A.\\
$^{3}$ Department of Biostatistics, Brown University, Providence, Rhode Island, U.S.A. \\
$^{4}$ Department of Data Science, Dana Farber Cancer Institute, Boston, Massachusetts, U.S.A. \\ 
$*$ \texttt{dafne$\_$zorzetto@brown.edu}\\}

\vspace{1cm}

\section*{A. Statistical Guarantees}

The proposed prior for treatment-specific factor scores offers several strengths, as outlined in Section~3.2. Additionally, it serves as a powerful tool for inferring information about the relationships among the multiple outcomes, as governed by unobserved features $U$. Specifically, the flexibility of the DDP distribution enables it to capture this unmeasured information while effectively controlling bias.

In particular, we consider the case where $U\sim \Norm(\mu_u, \sigma_u^2)$ with non-negative correlation $\sigma^{(lu)}$ between $U_i$ and $l_{it\review{h}}$, allowing us to derive closed-form expressions that clarify the relationships among the quantities involved.

Using multivariate normal properties,  
we can rewrite the marginal distribution of treatment-specific factor scores $\lt_{itj}$ and $U_i$, for each unit $i$, treatment $t$, and factor $j \in \{1, \dots, J_t\}$, as follows:
\begin{gather}
    \left[\begin{array}{c}
        \lt_{it\review{h}} \mid \x_i \\ U_i \end{array} \right]\sim \sum_{r \geq 1}\pi_{r}^{(t\review{h})}(\x_i) \Norm \left(\left[\begin{array}{c}
        \eta_r^{(t\review{h})} \\ \mu_u \end{array}\right], \Sigma_r^{(LU)} \right),
        \quad \Sigma_r^{(LU)} = \left[\begin{array}{cc}
        1/\tau_r^{(t\review{h})} & \sigma^{(lu)} \\ \sigma^{(lu)} & \sigma_u^2 \end{array}\right]. \label{eq:multinom}
\end{gather}
From the properties of the multivariate normal distribution, the conditional distribution of the factor scores given the covariates and the unmeasured variable is
\begin{equation}
    \{{\bf l}_{it\review{h}} \mid \x_i, U_i\} \sim \sum_{r \geq 1}\pi_{r}^{(t\review{h})}(\x_i)\Norm(\gamma_{0r}+\gamma_{1r}U_i, \sigma^{(l|u)2}_{r} ),\label{eq:multi_cond}
\end{equation}
where the location and scale parameters are
\begin{gather}
    \gamma_{0r}+\gamma_{1r}U_i = \eta_r^{(t\review{h})} + \frac{\sigma^{(lu)}}{\sigma_u^2 }(U_i-\mu_u),
   \quad \sigma^{(l|u)2}_{r} = \frac{1}{\tau_r^{(t\review{h})}}+ \frac{\sigma^{(lu)}}{\sigma_u^2 }. \notag
\end{gather}
Although the conditional distribution in Equation~\eqref{eq:multi_cond} cannot be directly estimated as $U$ is not measured, the following property \ref{prep:multinom_properties} demonstrates that our proposed formulation for the distribution of treatment-specific factors ${\bf L}_t$ is sufficient to recover the information about $U$ on average.

\begin{property}
\label{prep:multinom_properties} Assuming the unmeasured variable $U$ is normally distributed, the treatment-specific factors ${\bf L}_t$, for each $t$, are defined as in Eq.s (3)-(5) in Section 3, and the assumptions of Eq.\eqref{eq:multinom} hold, then each component of the mixture $r \geq 1$, has: (i) a finite expected value of the location parameter, that depend on the expected value of the unmeasured variable $U$, and (ii) the scale parameter has the scale parameter of the conditional distribution \eqref{eq:multi_cond} as a lower bound. Specifically,
\begin{align}
    &(i)\quad \E[\eta_k] = \gamma_{0r}+\gamma_{1r}\mu_u = \E[{\bf l}_{t\review{h}} \mid \x, U], \notag\\
    & (ii)\quad\frac{1}{\tau_r^{(t\review{h})}} = \sigma^{(l|u)2}_{r} + \frac{\sigma^{(lu)}}{\sigma_u^2 } \geq \sigma^{(l|u)2}_{r}. \label{eq:var_property}
\end{align}
Therefore, the probability distribution of $\{L_{t\review{h}}\mid \X\}$ has same mean as $\{L_{t\review{h}}\mid \X, U\}$ and an equal or larger variance. Substituting this result into Eq. (1), we conclude that the casual effects---the {\bf SATE}---are unbiased under the assumptions defined in Section~2.2.
\end{property}

Scenario 1 in Figure~1 corresponds to the case where factor scores and the unmeasured variable are independent, i.e., $\sigma^{(lu)}= 0$. In this case, the marginal and conditional variances are the same, indeed the equation \eqref{eq:var_property} can be simplified in $1/\tau_r^{(t\review{h})} = \sigma^{(l|u)2}_{r}$. Conversely, Scenarios 2 and 3 assume $\sigma^{(lu)} > 0$. 

\bigskip

\section*{B. Posterior Computations}
We implement a Gibbs sampling algorithm for model fitting, outlined below, which builds on the algorithms proposed by \cite{bhattacharya2011sparse} for the estimation of factor loading and \citet{rodriguez2011nonparametric} for the prior of the . 

Following the steps of the algorithm~\ref{alg:gibbs}, in each iteration $r=1,\dots, R$, we use the observed data $(y,t,\x)$ to update the parameters and latent factors and then impute the missing potential outcomes $y^{mis}$.
Given initial random values for parameters and latent variables, we recurrently draw them in the following order.

\vspace{0.25cm}
\noindent {\em Covariates regression.} The Gaussian conjugate prior distribution for the covariates regression parameter allow us to write the posterior distribution for them as following
\begin{gather*}
    \{\B_{tk} \mid \x, \y, \Lambda, \lt, \mu_\beta, \sigma_\beta\} \sim \Norm(V_\beta^{-1} M_\beta, V_\beta^{-1}), \\
    V_\beta = \sigma_\beta^{-1}\iv_p + \psi_{tk}^{-1}\x^T \x, \\
    M_\beta = \mu_\beta/ \sigma_\beta\iv_p + \psi_{tk}^{-1}(\x^Ty_{k} - \x_i^T\Lambda_{t[k,\cdot]}\lt_{tk});
\end{gather*}
for each treatment level $t \in \{0,1\}$ and each component of the outcome variable $k=\{1, \dots, q\}$.

\vspace{0.25cm}
\noindent {\em Factor scores.} The posterior distribution for the treatment-specific factor score conditional to the cluster allocation ${\bf S}_{it}$ for each unit $i \in \{1, \dots, n\}$ is 

\begin{gather*}
    \{ \lt_{it} \mid {\bf S}_{it} = \Bar{\bf s}, \x, \y, \Lambda, \lt, \mu_l, \tau \} \sim \Norm(V_{l\mid s}^{-1} M_{l\mid s}, V_{l\mid s}^{-1}) \\
    V_{l\mid s} = \left(\mathrm{T}^{(t)}_{\Bar{\bf s}}\right)^{-1} + \Lambda_{t}^{T} \Psi_{t}^{-1} \Lambda_{t} \\
    M_{l\mid s} = \Lambda_{t}^{T} \Psi_{t}^{-1} (\y_i-\B_t\x_i) + \mu^{(t)}_{ \Bar{\bf s}}\left(\mathrm{T}^{(t)}_{\Bar{\bf s}}\right)^{-1}
\end{gather*}
where ${\bf S}_{it}$ is the vector of cluster allocation latent variable of each $\lt_{it}$ for the unit $i$ and treatment $t$, such that $\Bar{\bf s}=\{s_{it1}, \dots, s_{it k}\}$ are the current values.
The $\mu^{(t)}_{ \Bar{\bf s}}= \{\mu^{(t)}_{s_{it1}}, \dots, \mu^{(t)}_{s_{itk}} \}$ and $\mathrm{T}^{(t)}_{ \Bar{\bf s}}=\mbox{diag}(\tau^{(t)}_{s_{it1}}, \dots , \tau^{(t)}_{s_{itk}})$ are respectively the means and variances of the $\Bar{\bf s}$ components of the mixture.

\vspace{0.25cm}
\noindent {\em Factor loading.} If we denote the $j$-th row in $\Lambda_t$ by $\lambda_{jt}^{\top}$ for each $t\in\{0,1\}$. The conditional distributions for $\lambda_{jt}$ is the following 
\[
\{\lambda_{jt} \mid \cdots\} \sim \Norm \left(
\left( D_{jt}^{-1} + \psi_{jt}^{-2} l_t^{\top} l_t \right)^{-1}
 l_t^{\top} \psi_{jt}^{-2} \tilde{y}_{tj},
\left( D_{jt}^{-1} + \psi_{jt}^{-2} l_t^{\top} l_t \right)^{-1}
\right),
\]
where $\lt_t = (\lt_{1t}, \ldots, \lt_{n_t t})^{\top}$, $\tilde{y}_{tj} = (y_{1tj}-\X_1\beta_t^T, \ldots, y_{n_t tj}-\X_{n_t}\beta_t^T)$ indicate the observed outcome removing the regression given the current value for $\boldsymbol{\beta}_t$, and 
$D_{jt}^{-1} = \mathrm{diag}(\theta_{jt1}^{t} \iota_1^{t}, \ldots, \theta_{jtJ_t}^{t} \iota_{J_t}^{t})$ is function of the parameters of the \citet{bhattacharya2011sparse}'s prior.

\vspace{0.25cm}
\noindent {\em Hyperparameters for factor loading.} The posterior distribution of $\theta_{tjh}$ and $\delta_{tl}$, for each factor $j$, component $p$ of the outcome, and treatment $t$, are respectively
\begin{align*}
    & \{\theta_{tjh} \mid -\} \sim \Gamma\left( \frac{\nu_t + 1}{2}, \frac{\nu_t + \iota_{th} \lambda_{tjh}^2}{2} \right),\\
    & \{\delta_{t1} \mid -\} \sim \Gamma\left(
a_{1_t} + \frac{p k}{2},\ 
1 + \frac{1}{2} \sum_{j=1}^{k} \iota_{tj} \sum_{h=1}^{p} \theta_{tjh} \lambda^2_{tjh}
\right), \\
& \{\delta_{tl} \mid -\} \sim \Gamma\left(
a_{2_t} + \frac{p}{2}(k - l + 1),\ 
1 + \frac{1}{2} \sum_{j=l}^{k} \iota_{tj} \sum_{h=1}^{p} \theta_{tjh} \lambda^2_{tjh}
\right), \quad  l \geq 2.
\end{align*}

\vspace{0.25cm}
\noindent {\em Cluster allocation.} The latent variables $S_{itj}$ identifies the cluster allocation for each units $i$ at the treatment level $t$ and factor $j=\{1, \dots, k\}$. Its posterior distribution is a multinomial distribution where
\begin{equation*}
    \mathbb{P}(S_{itk}=l \mid \eta_l^{(tj)},\tau_l^{(tj)} ) \propto \pi_{l}^{(tj)}(\x_i)\Norm(\lt_{it};\eta_l^{(tj)}, 1/\tau_l^{(tj)}),
\end{equation*}
for $l=1,\dots,L$, with $\pi^{(tj)}_{l}$ defined as:
\begin{equation*}
    \pi_{l}^{(tj)}(\x_i)=\Phi(\alpha_{l}^{(tj)}(\x_i)) \prod_{r<l} (1-\Phi(\alpha_{l}^{(tj)}(\x_i))),
\end{equation*} 
for $l=1,\dots,L-1$ and with $\Phi(\alpha_{L}^{(tj)}(\x_i))=1$.

\vspace{0.25cm}
\noindent {\em Cluster Specific Parameters.}
Thanks to the draw values for the latent variables $S_{itj}$, we know for each cluster $l \in \{1,\dots, L\}$, the allocated units and we can update the values of the parameters from their posterior distributions:
\begin{align*}
    \eta_l^{(tj)} & \sim  \Norm\left( V_\eta^{-1}\times \left(\frac{\sum_{\{i: S_{itk}=l\}}\lt_{it}}{\sigma_{1}^{(t)2}} +\mu_\eta\tau_l^{(tj)} \right),V_\eta^{-1}\right);\\
    \tau_l^{(tj)} & \sim \mbox{Gamma}\left(\gamma_1+ \frac{n_{l}^{(t)}}{2},\gamma_2+\frac{\sum_{\{i: S_{itk}=l\}} \big[\lt_{it}-\eta_l^{(tj)}\big]^2}{2}\right), \mbox{ for } l=1,\dots,L;
\end{align*}
where $V_\eta=n_{l}^{(tj)}\tau_l^{(tj)} +1/\sigma^2_{\eta}$ and $n_{l}^{(tj)}$ is the number of units allocated in the $l$-th cluster.

\vspace{0.25cm}
\noindent {\em Augmentation Scheme.}
In order to sample from $\{a_{l}^{(tj)}(\x)\}_{l=1}^L$ and the corresponding weights $\{ \pi_{r}^{(tj)}(\x)\}_{l=1}^L$, we need a data augmentation scheme. The idea was developed by Albert and Chib (2001).  and borrowed by \citet{rodriguez2011nonparametric} to obtain exact Bayesian inference for binary regression and computationally easy to include it in the Gibbs sampling (Albert and Chib, 2001). We can impute the augmented variables $Z_{l}^{(tj)}(\x_i)$ by sampling from its full conditional distribution \citep{rodriguez2011nonparametric}:
\[
    Z_{l}^{(tj)}(\x_i)|S_{i}^{(t)},\alpha_{l}^{(t)}(\x_i) \sim \begin{cases}
    \Norm(a_{l}^{(tj)}(\x_i),1)\iv_{\R^+} \mbox{ if } S_{itk}=l,\\
    \Norm(a_{l}^{(tj)}(\x_i),1)\iv_{\R^-} \mbox{ if } S_{itk}<l.
\end{cases}
\]
The mean, $a_{l}^{(tj)}(\x_i)$, of the previous normal distributions is obtained from:
\[
    a_{l}^{(tj)}(\x_i)=\phi\left(\frac{\pi_{l}^{(tj)}(\x_i)}{\prod_{r<l}(1-\Phi(\x_i^T\betabold_{l}^{(t)})}\right)= \phi\left(\frac{\pi_{r}^{(tj)}(\x_i)}{1-\sum_{r<l}\pi_{r}^{(tj)}(\x_i)}\right);
\]
where $\phi(\cdot)$ is the continuous density function of Gaussian distribution. 

\vspace{0.25cm}
\noindent {\em Weights in the mixture}
To conclude the for-loop, the $\{\alpha_{ql}^{(tj)}\}_{q=0}^p =(\alpha_{0l}^{(tj)},\betabold_{l}^{(tj)})$, for  $l=1,\dots,\max(S_{itk},L-1)$, are updated for the posterior distribution:
\begin{align*}
    \alpha_{0l}^{(t)} & \sim \Norm( (1/\sigma^2_{\beta}+n)^{-1}\times (\mu_\beta/\sigma_\beta^2 +1_n^T\Tilde{\bf Z}),(1/\sigma^2_{\beta}+n)^{-1}); \\
    \betabold_{l}^{(t)} & \sim \Norm_{p}( W^{-1}\times (\mu_\beta/\sigma_\beta^2 +(\Tilde{\X})^T\Tilde{\bf Z}),W^{-1});
\end{align*}
where $1_n$ is a $n$ vector of ones, $W=I_{p}/\sigma^2_{\alpha}+(\Tilde{\X})^T\Tilde{\X}$, $I_{p}$ is a  $p\times p$ diagonal matrix, $\Tilde{\X}$ is a matrix such that it is composed by the rows $i$ in $\X$, such that $S_{itk}\leq l$, and $\Tilde{\bf Z}$ is a vector composed by the $z_{l}^{(tj)}(\x_i)$ for the units $i$ and factor $j$ such that $S_{itk}\leq l$.

{\vspace{0.8cm}
\begin{algorithm}
\caption{Estimation Multivariate Causal Effect with Factor Model}\label{alg:gibbs}
\vspace{0.15cm}
{\bf Inputs:} 

\quad - the observed data $({\bf y},t,{\bf x})$;

\quad - upper bound for the number of clusters for the PSB truncation.

{\bf Outputs:} 

\quad - posterior distributions of parameters: $\{{\bf B}_t\}_{t=(0,1)}$;

\quad -  posterior distributions of factors:$\{\Lambda_t\}_{t=(0,1)}$ and $\{\lt\}_{t=(0,1)}$;

\quad -  distributions of potential outcome: ${\bf Y}(0), {\bf Y}(1)$.

\vspace{0.11cm}

{\bf Procedure:}
\begin{algorithmic}
\State Initialization of all parameters and latent variables.
\For{$r \in \{1,\dots,R\}$}
\For{$t\in \{0,1\}$}
    \State Draw the covariates-regression parameters $\beta_t$;
    \State Draw the treatment-specific factor scores given cluster allocation $\{\lt_{ti} \mid {\bf S_{it}}\}_{t=(0,1), \forall i}$;
    \State Draw the treatment-specific factor loadings $\Lambda_t$;
    \State Draw the parameters $\kappa_t$ and $\iota_t$ for treatment-specific factor loadings;
    \State Compute the cluster allocation for the treatment-specific latent factor scores $\{{\bf S_{it}}\}_{t=(0,1), \forall i}$;
    \State Draw parameters of the Gaussian distributions in the mixture;
    \State Augmentation scheme for the probit stick-breaking process;
    \State Compute weights in the mixture $\boldsymbol{\pi}$;
    \State Imputation the potential multivariate outcomes $\{ {\bf Y}_i(0), {\bf Y}_i(1)\}_{\forall i}$.
\EndFor
\EndFor
\end{algorithmic}
\end{algorithm}}

\bigskip

\section*{C. Further Details of Simulation Study}
\bigskip \section*{\small Simulation study details}

In the main text, we reported four simulation scenarios. The first three explore different relationships between the unmeasured variable $U$ and the confounders $\X$, corresponding to the graphical representations in Figure~1 in the main text. 
The fourth scenario is designed to closely mimic the real dataset used in our application in Section~5. Following the details  provided in Table~1 in the main text, we have the following simulation-generating processes.

In the first three scenarios, the treatment-specific factor scores $\{{\bf L}_0, {\bf L}_1\}$ depend on both the confounders $\X$ and the unmeasured variable $U$, while the potential outcomes $\Y$ depend on $U$ only through the factors. In Scenario~1, shown on the left of Figure~1, 
$\X$ and $U$ are independent (i.e.,  $\sigma^{lu}=0$ in Property~1 in Section A, Supplementary Materials). Scenario 2 assumes that $U$ depends on $\X$, and Scenario 3 explores the reverse relationship, where the confounders $\X$ depend on $U$. These dependencies are highlighted by the red dotted arrows in the second and third representations of Figure~1. 
Accordingly to Scenario 2, we consider four independent confounders $\X$ and model $U$ as dependent on $\X$. In contrast, Scenario 3 is modeled with an independent normal distribution for $U$; also the first two confounders in $\X$ are normally distributed with mean given by a linear function of $U$.
Scenario 4 uses real-data information for the 28 confounders and the treatment assignment. Factor loadings are estimated with our model applied to real data. Factor scores and outcomes were simulated according to our proposed model.

\bigskip \section*{\small Supplementary scenarios}

\review{Additionally to the simulation study included in the main text we simulated two more scenarios: Scenario 5 analyzes the case when the outcomes are not highly correlated and the mean function $g_t(X)$ is non-linear, while Scenario 6 consider the case when the unmeasured variable $U$ directly affect the outcomes.}

\review{Specifically, Scenario~5 follows the same general specification as Scenario~1, as summarized in Table~1 of the main text, with two modifications. First, the factor loading matrix $\Lambda_t$ for $t = 0,1$ has entries $\lambda_{th}$ drawn from a mixture of two Uniform distributions, with support on $[-0.15, -0.05]$ and $[0.05, 0.15]$. This construction induces correlations among the outcomes that are close to zero. Second, the mean functions $g_t(X)$ are nonlinear and are defined separately for each treatment level as follows:}
\begin{gather*}
    g_t({\bf X}) = \beta_{t0}+ \beta_{t1}e^{X_1} + \beta_{t2}X_2^2 + \beta_{t3} X_3+ \beta_{t4}X_4 + \beta_{t5}\mathbb{I}_{(X_3=1, X_4=1)}+ \beta_{t6}\mathbb{I}_{(X_3=0, X_4=0)}.
\end{gather*}

\review{Scenario~6 follows the same general specification as Scenario~1, as summarized in Table~1 of the main text, with the difference that the regression of the potential outcome is linear correlated with the unmeasured variable $U$, such that $Y(t) = B_t{\bf X} + \beta_{tu}U + \Lambda_t l_{t} +\xi_t$. }

\review{As reported in Figure~\ref{fig:sim_appendix}, Scenario~5 show similar results across model for the bias and MSE, with exception of \textit{StandardFA}, that in half of the outcome variables does not include the zero in the interquantile range visualized in the box of the bias boxplots. Moreover, the coverage rate in Figure~\ref{fig:coverage} show a notable undercoverage for the competitor model \textit{StandardFA}. While all the model in Scenario~6 are biased, with smaller MSE values for the non-parametric models {\it BART} and {\it BCF}, and a coverage equal to zero. The results for this final scenario are expected, as none of the considered models is designed to account for unmeasured confounding and all rely on the ignorability assumption. }

\begin{stdfigure}[h!]
\centerline{
    \includegraphics[trim={0cm 14cm 0cm 0cm},clip,width=190mm]{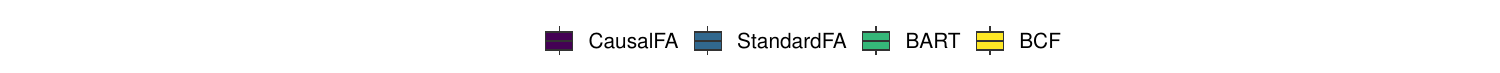}}
\centerline{
    \includegraphics[trim={0cm 0cm 0cm 9.5mm},clip,width=75mm]{plots/2nd_submission_sim/Scenario_5_Bias.pdf}
    \includegraphics[trim={0cm 0cm 0cm 9.5mm},clip,width=75mm]{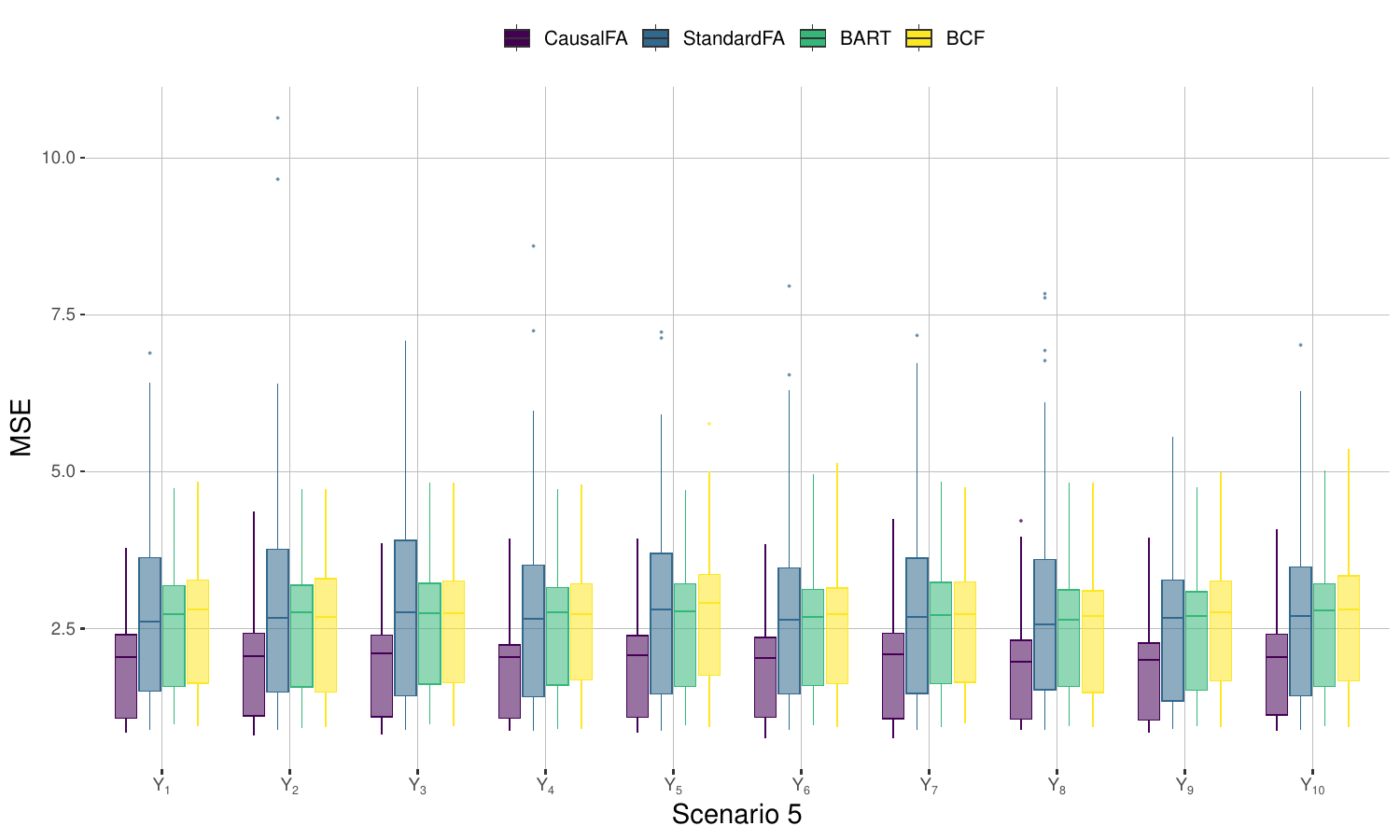}}
\centerline{
    \includegraphics[trim={0cm 0cm 0cm 9.5mm},clip,width=75mm]{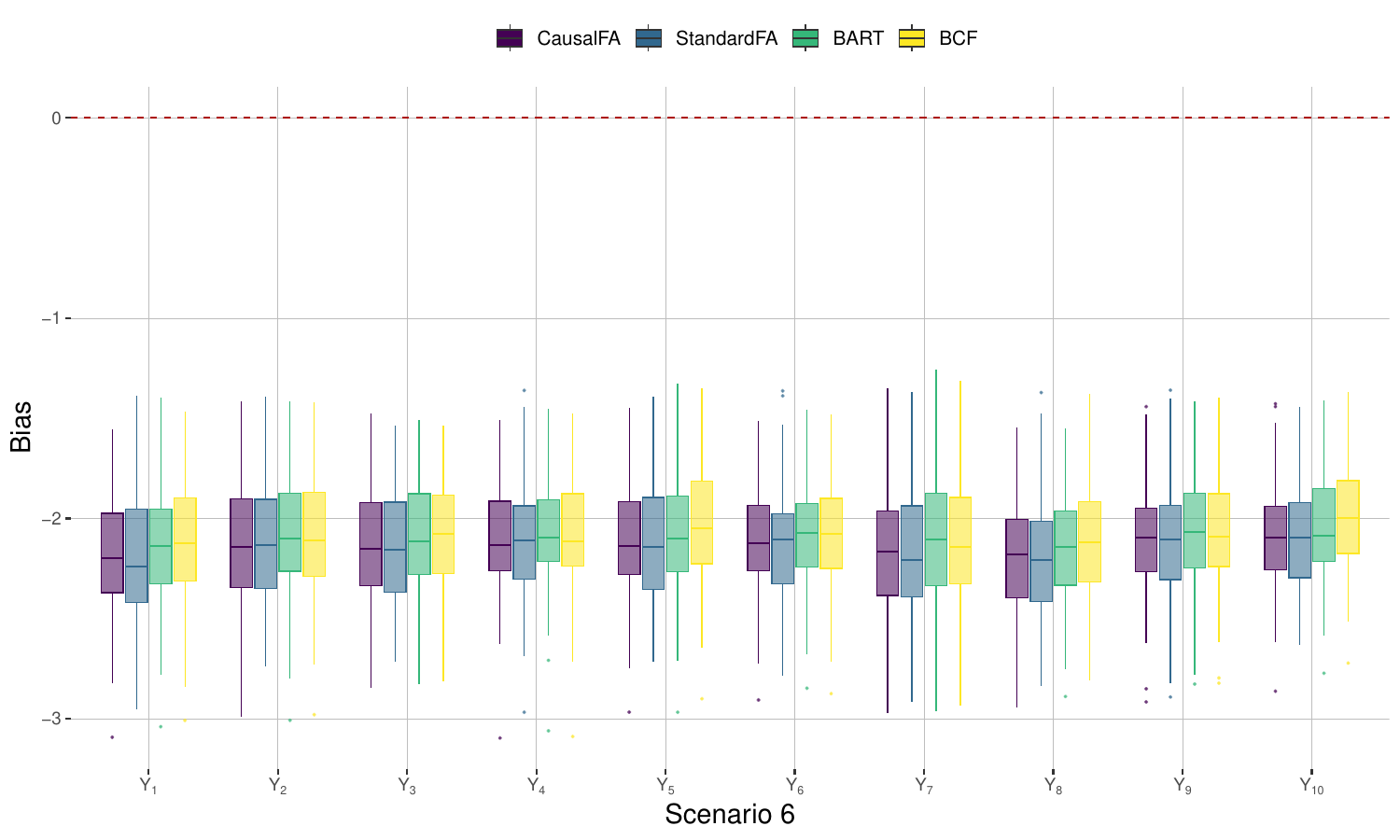}
    \includegraphics[trim={0cm 0cm 0cm 9.5mm},clip,width=75mm]{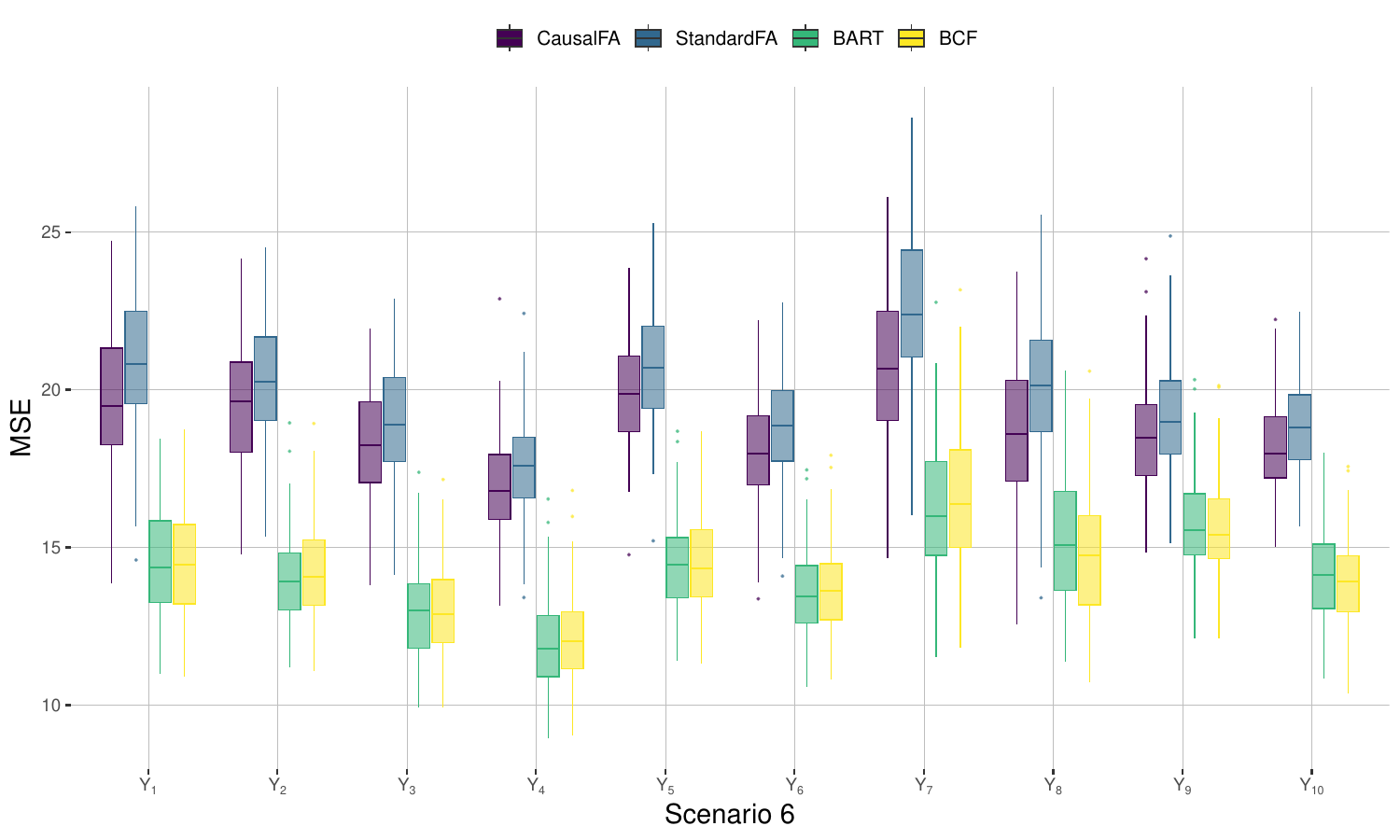}}
\caption{Bias (left) and mean square error (MSE, right) across simulated scenarios. Results are shown for our proposed model, standard factor model, causal BART, and BCF. \label{fig:sim_appendix}}
\end{stdfigure}

\review{We also compare the empirical coverage rates across all simulated scenarios for the four methods. Figure~\ref{fig:coverage} reports the 95\% coverage, with the nominal target value of 0.95 indicated by the dashed red line. In the first three scenarios, the intervals of our proposed model {\it CausalFA} include the target values, whereas \textit{StandardFA} exhibits undercoverage and both \textit{BART} and \textit{BCF} tend to overcover. In Scenario~4, \textit{BART} shows undercoverage, which, consistently with the corresponding bias and MSE results, indicates degraded performance in more complex data-generating settings.}

\begin{stdfigure}[h!]
\centerline{
    \includegraphics[trim={0cm 14cm 0cm 0cm},clip,width=190mm]{plots/2nd_submission_sim/Scenario_1_Bias.pdf}}
\centerline{
    \includegraphics[trim={0cm 0cm 0cm 9.5mm},clip,width=75mm]{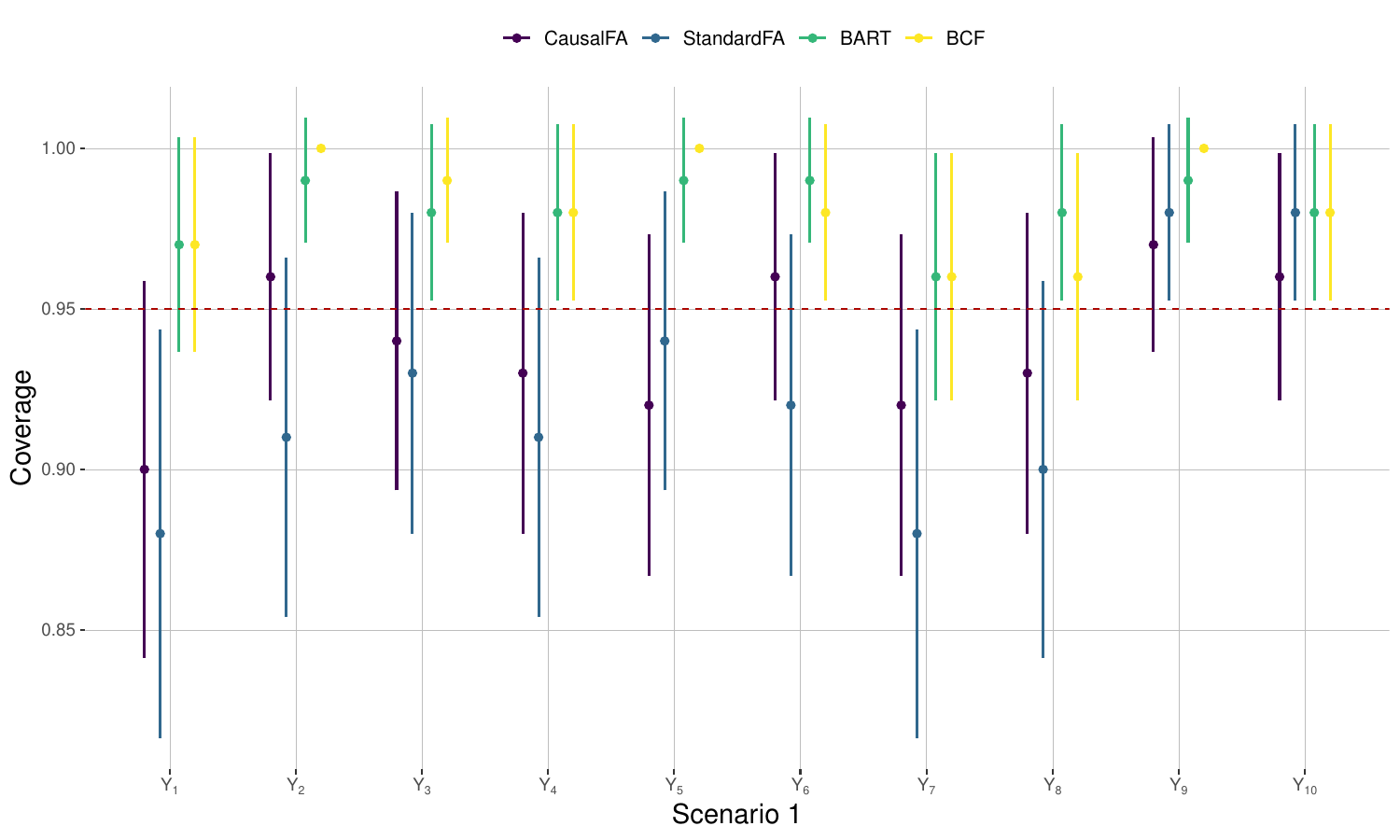}
    \includegraphics[trim={0cm 0cm 0cm 9.5mm},clip,width=75mm]{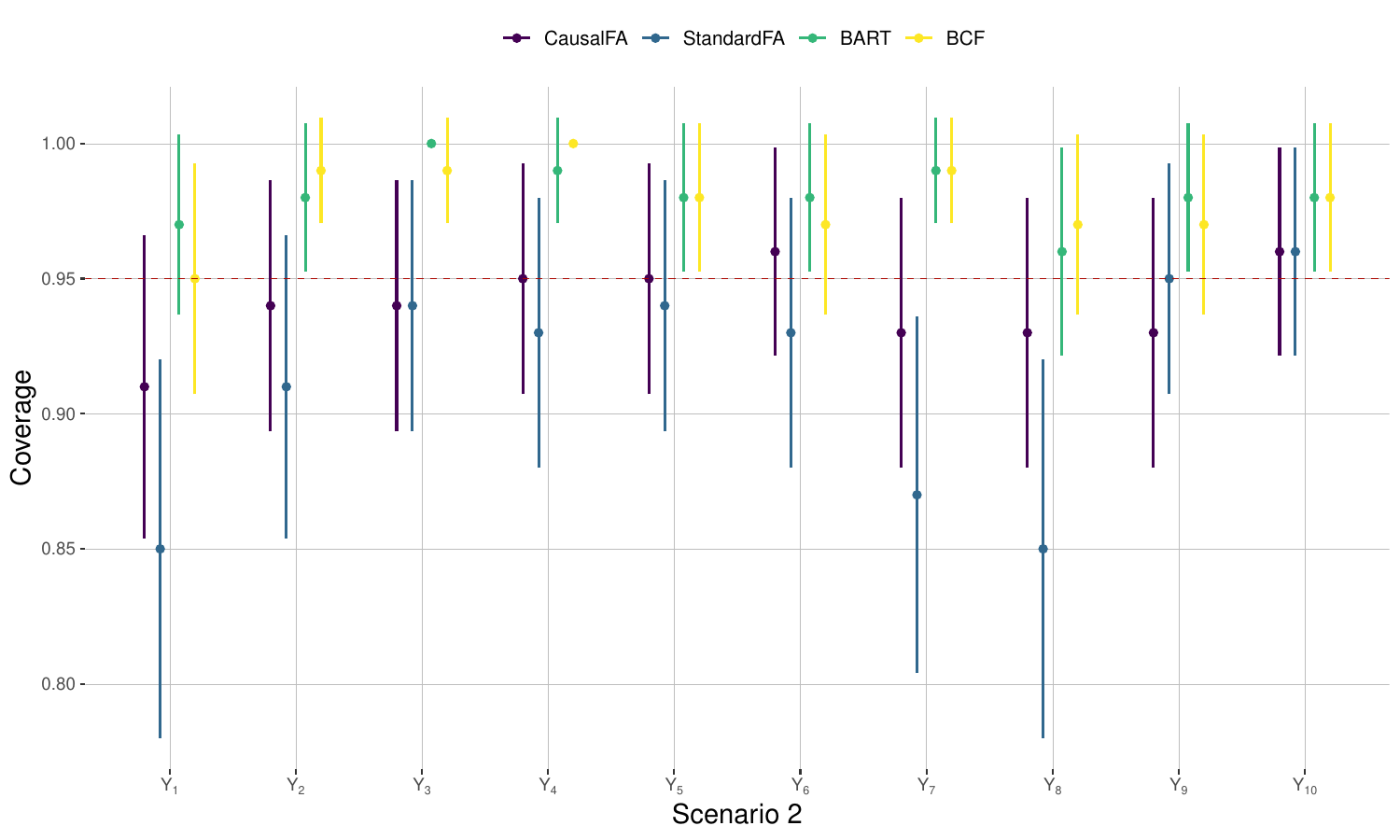}}
\centerline{
    \includegraphics[trim={0cm 0cm 0cm 9.5mm},clip,width=75mm]{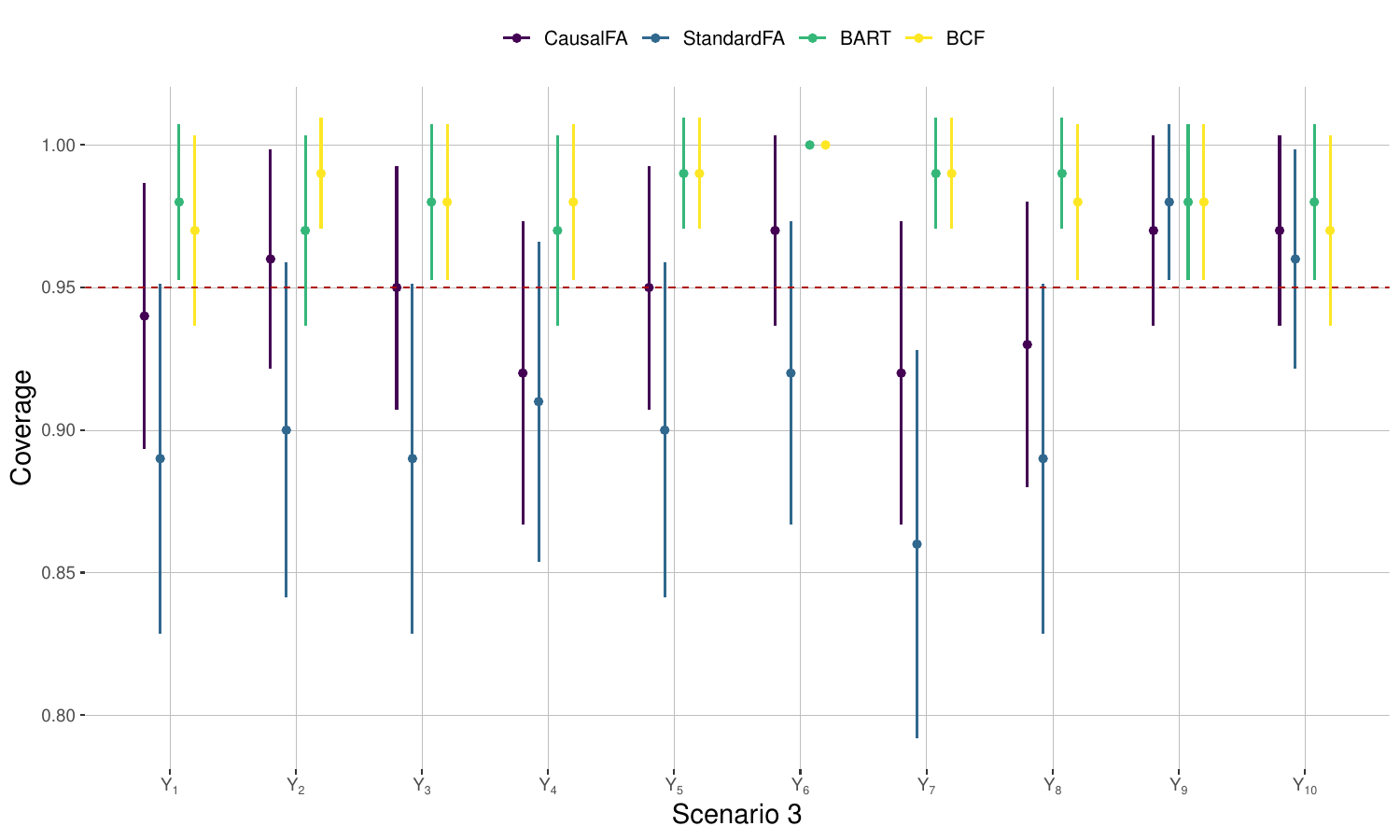}
    \includegraphics[trim={0cm 0cm 0cm 9.5mm},clip,width=75mm]{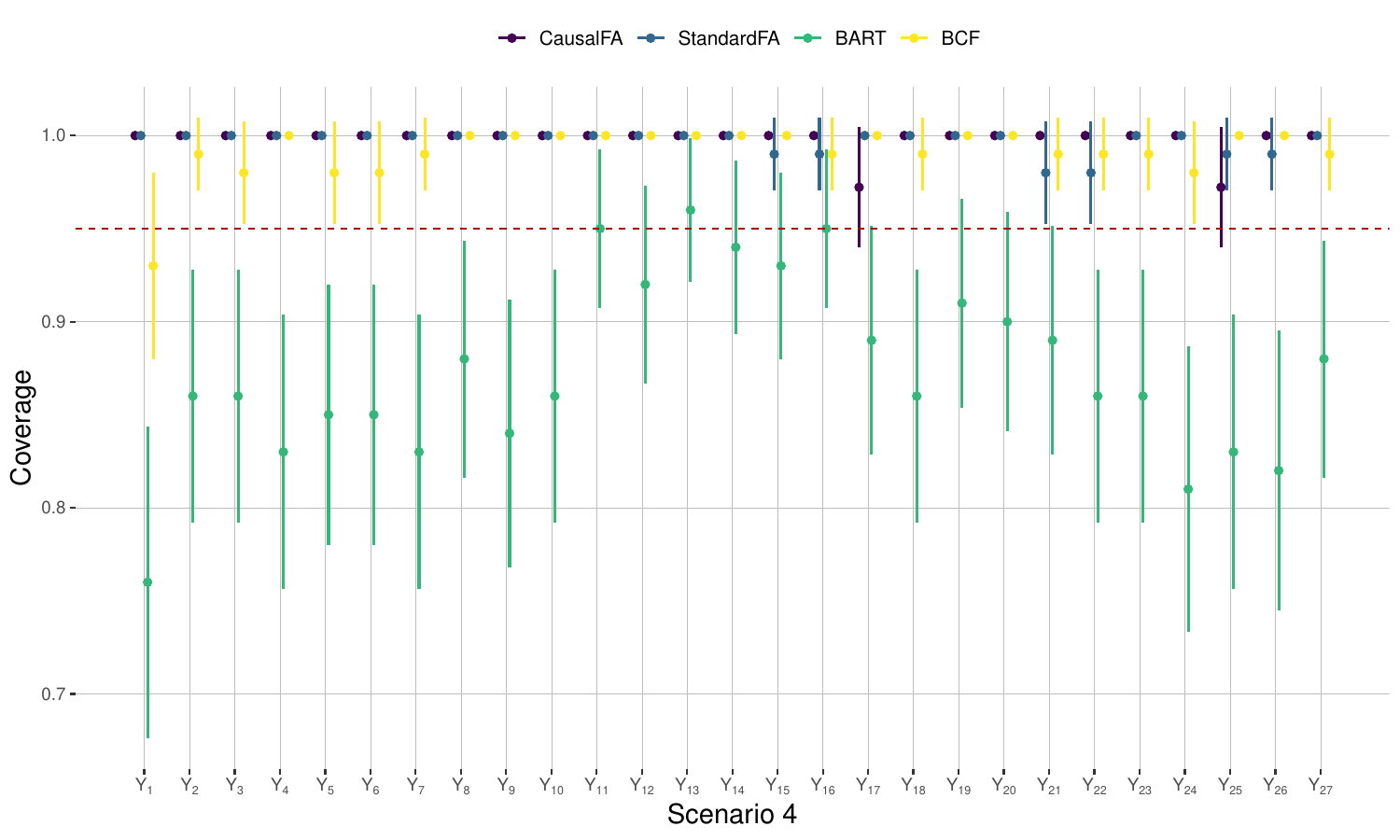}}
\centerline{
    \includegraphics[trim={0cm 0cm 0cm 9.5mm},clip,width=75mm]{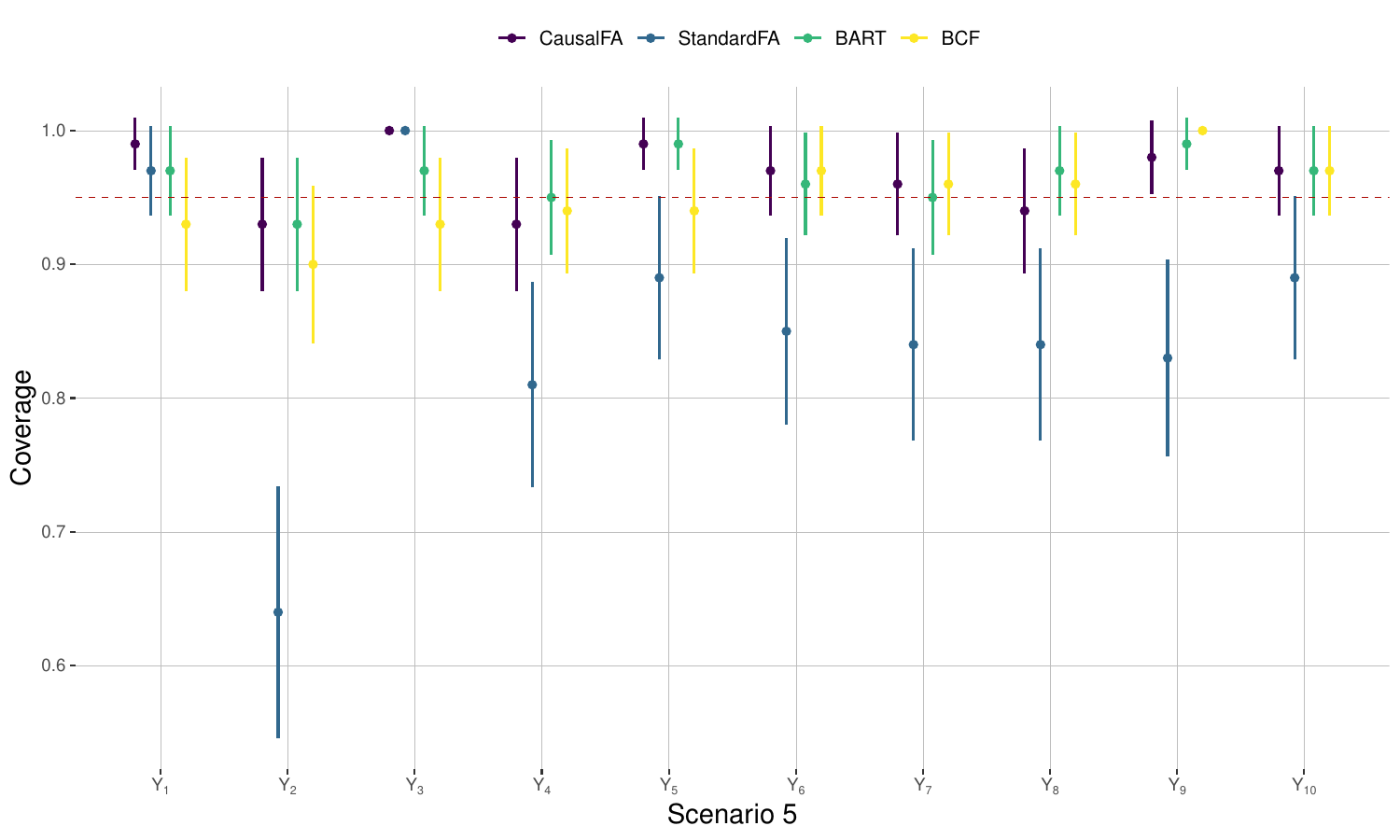}
    \includegraphics[trim={0cm 0cm 0cm 9.5mm},clip,width=75mm]{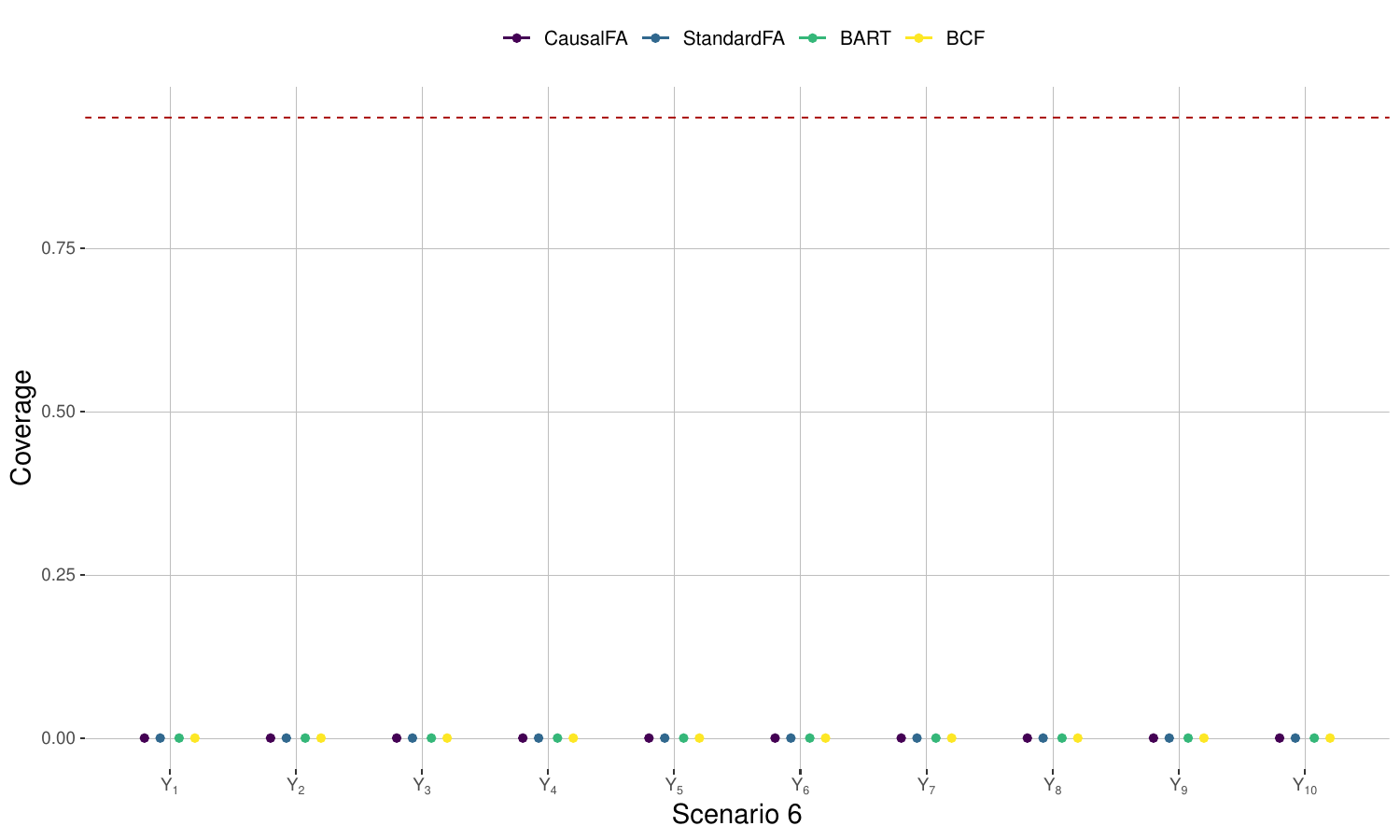}}
\caption{95\% coverage across simulated scenarios. Results are shown for our proposed model, standard factor model, causal BART, and BCF. The red horizontal line indicates the target value, equal to 0.95. \label{fig:coverage}}
\end{stdfigure}

\bigskip

\section*{D. Further Details of Application Dataset}
\bigskip \section*{\small Details of chemical species}

In our analysis, we select the same $27$ chemical species considered in \citet{krasovich2025influence}, divided in alkaline-earth metals (Magnesium, Calcium, and Strontium), alkali metals (Sodium, Rubidium, and Potassium),
transition metals (Chromium, Nickel, Vanadium, Copper, Iron, Zinc, and Manganese), metalloids (Arsenic and Silicon), other metals (Lead, Aluminum, and Titanium), nonmetals (Selenium, Nitrate, Sulfate, Sulfur, and Phosphorus), halogens (Chlorine and Bromine), and Organics (elemental Carbon and organic Carbon).

\bigskip \section*{\small Monitors for chemical species}

\review{In Section~5, we study the causal effects of wildfire smoke on chemical species in \PMns. The monitors measuring the concentrations of the chemical species are distributed across the continental United States; their locations are shown in Figure~\ref{fig:map}, which illustrates the spatial distribution of the monitoring sites across the study region.}

\begin{stdfigure}[h!]
\centering
\includegraphics[width=12cm]{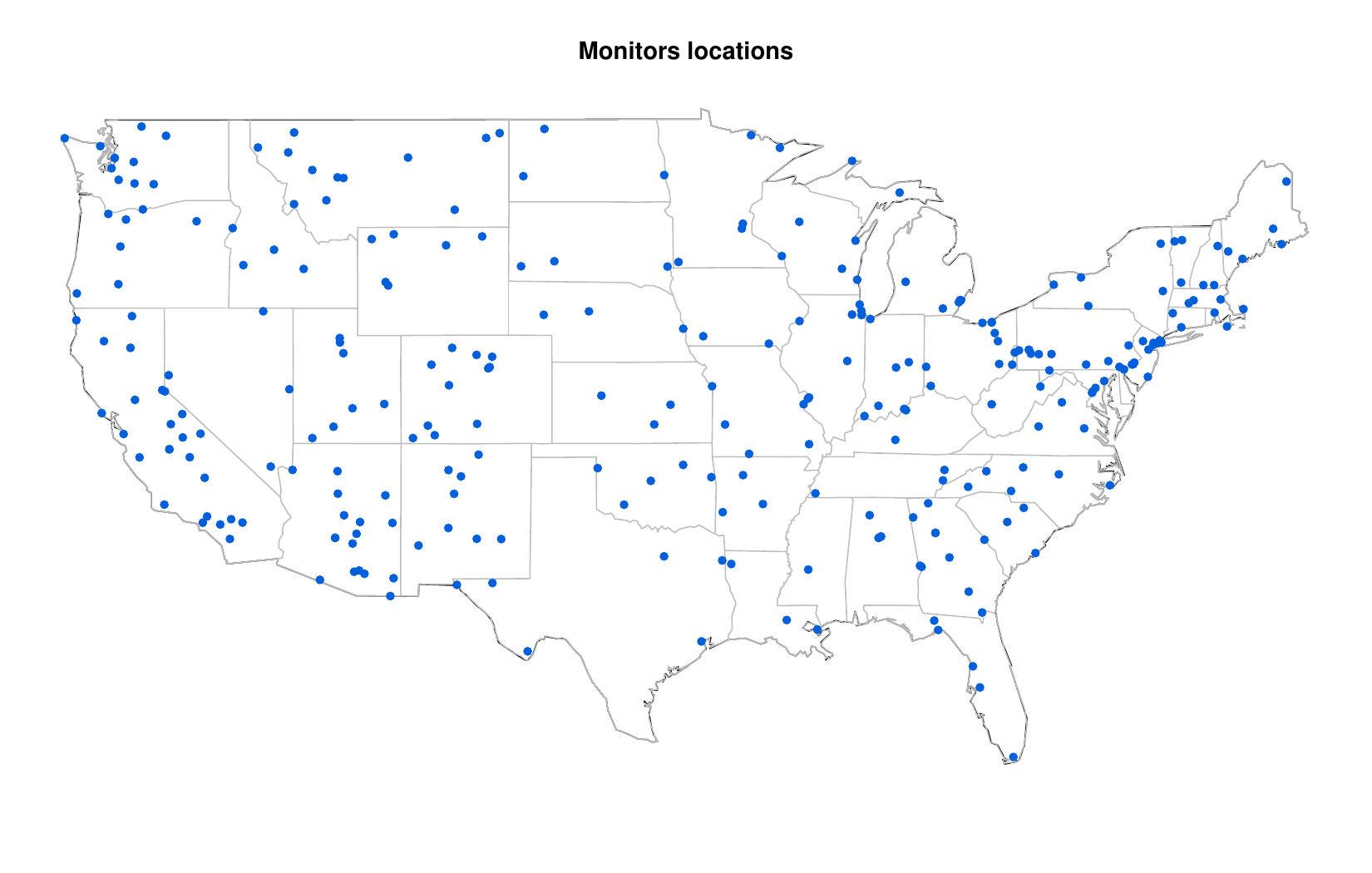} 
\caption{Map showing the spatial distribution of monitoring locations in the study dataset.\label{fig:map}}
\end{stdfigure} 

\review{Due to the nature of wildfire smoke, its occurrence varies in both frequency and duration across monitoring sites. Figure~\ref{fig:time_series} illustrates examples of the observed data. For the six monitors shown, the vertical pink lines indicate days with wildfire smoke at each location. The frequency and duration of smoke events differ notably across sites: no smoke days are observed at SIPS1, a few sporadic smoke days occur at CABI1 and 482030002, while SAMA1 and EVER1 experience many smoke days, including several consecutive events. }

\begin{stdfigure}[h!]
\centering
\includegraphics[width=7cm]{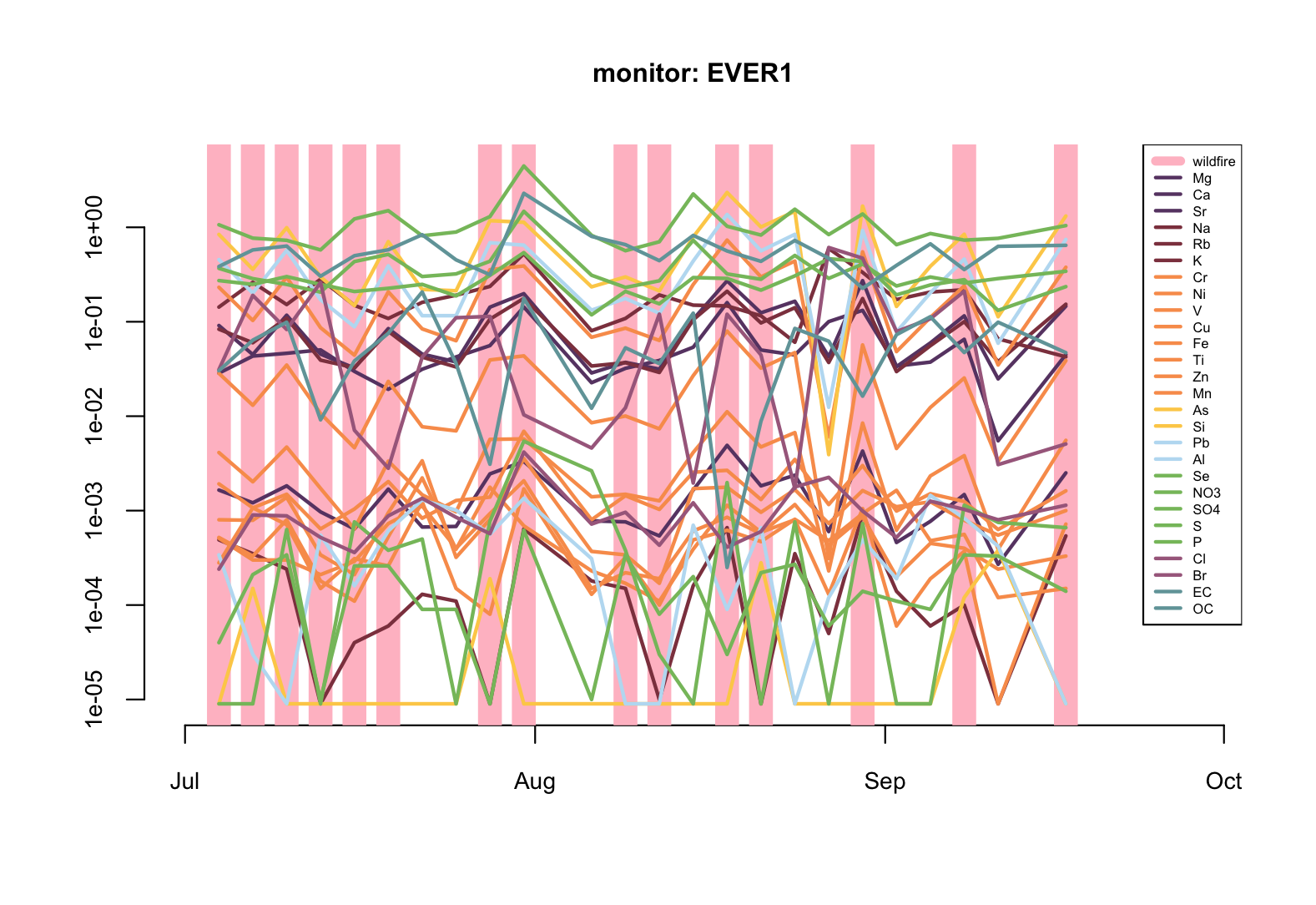}\; \includegraphics[width=7cm]{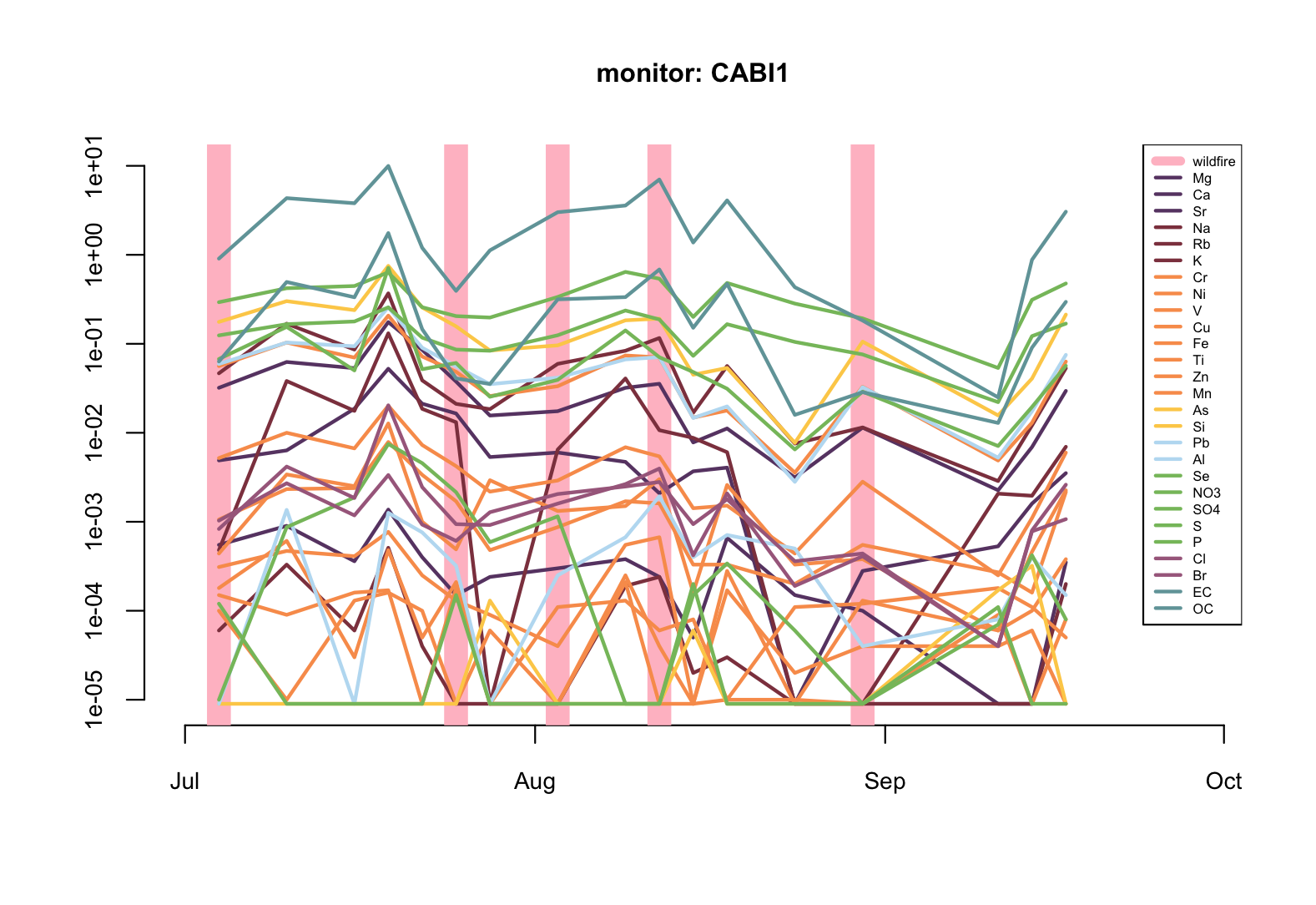} \\
\includegraphics[width=7cm]{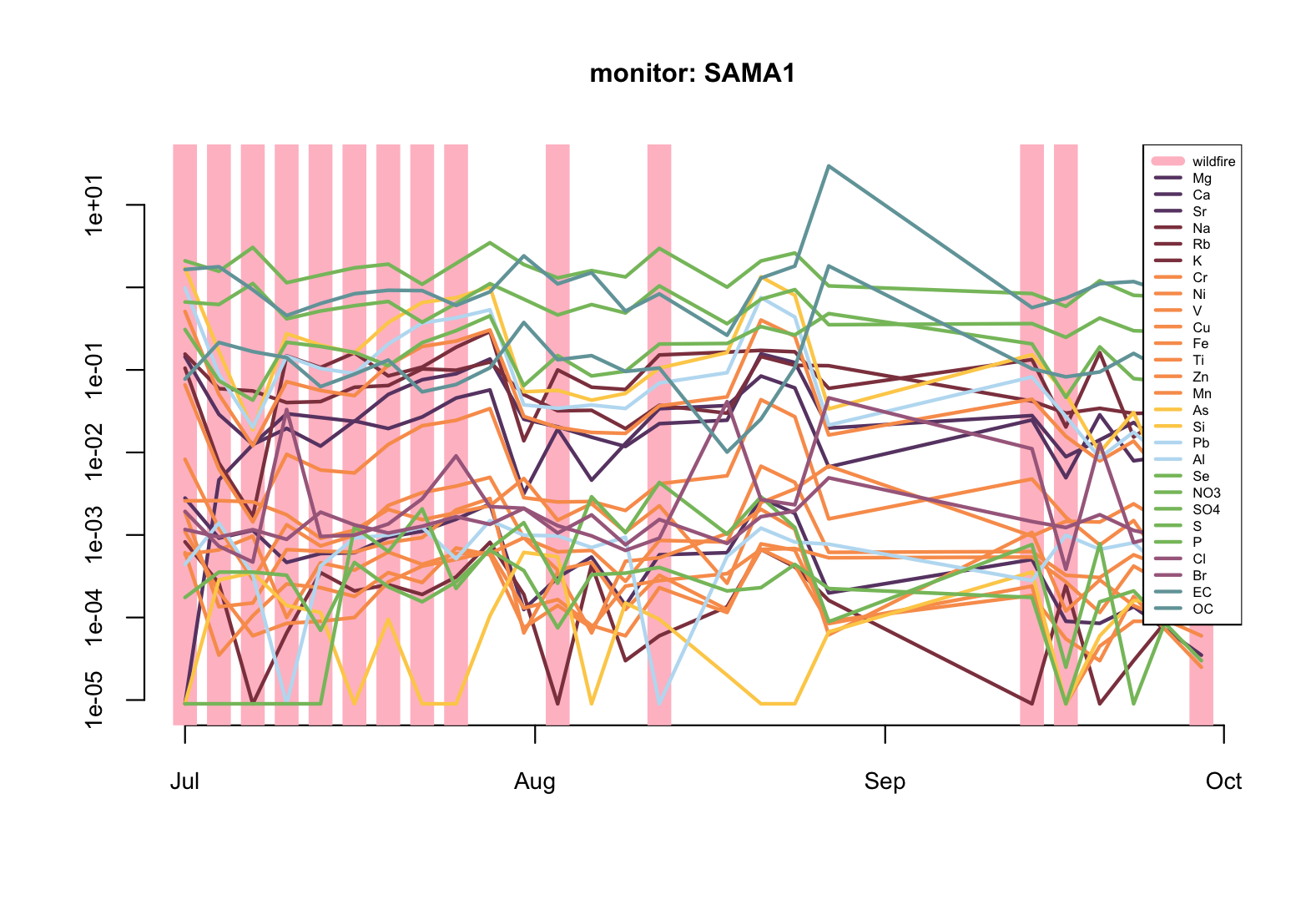} \;
\includegraphics[width=7cm]{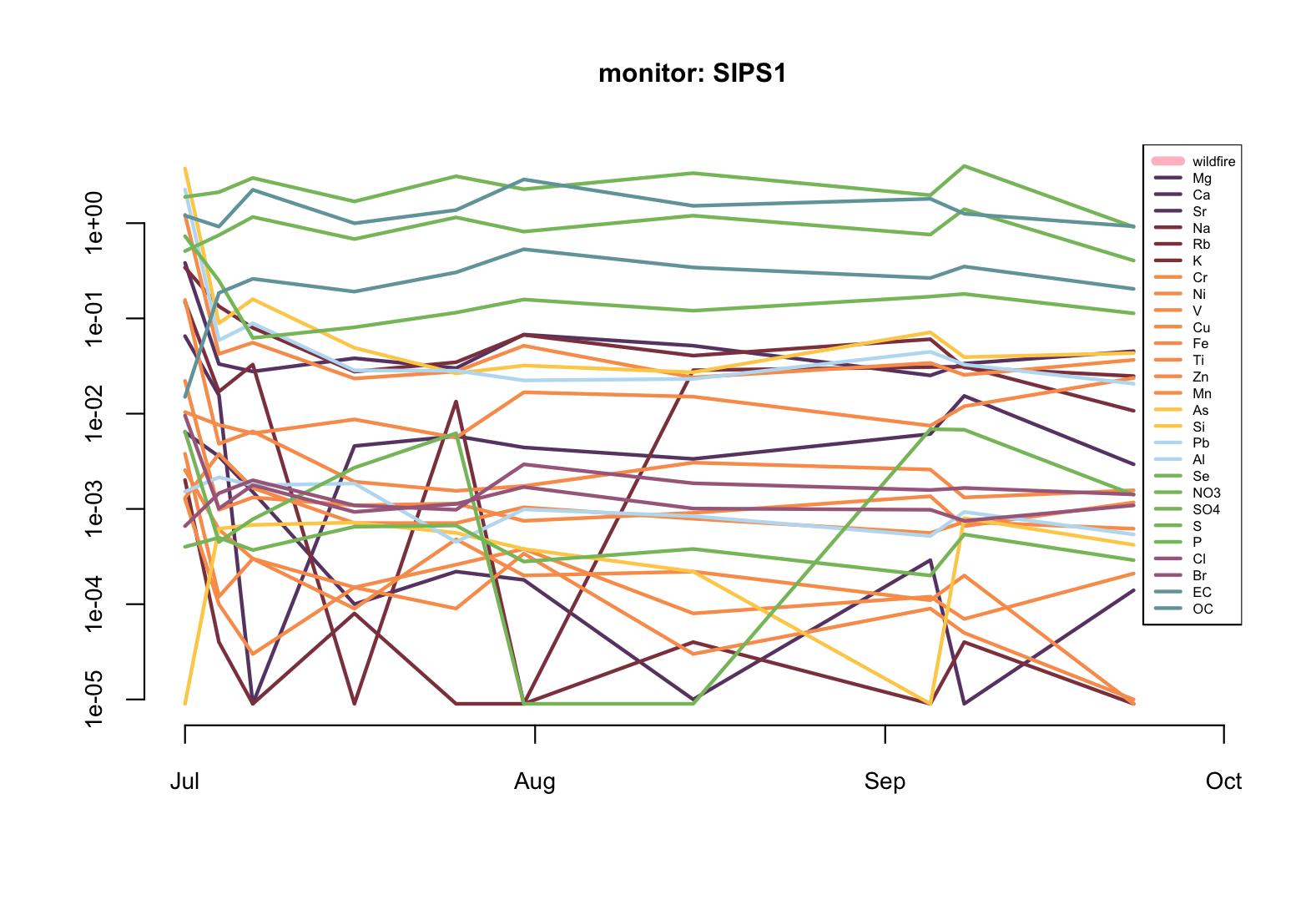} \\
\includegraphics[width=7cm]{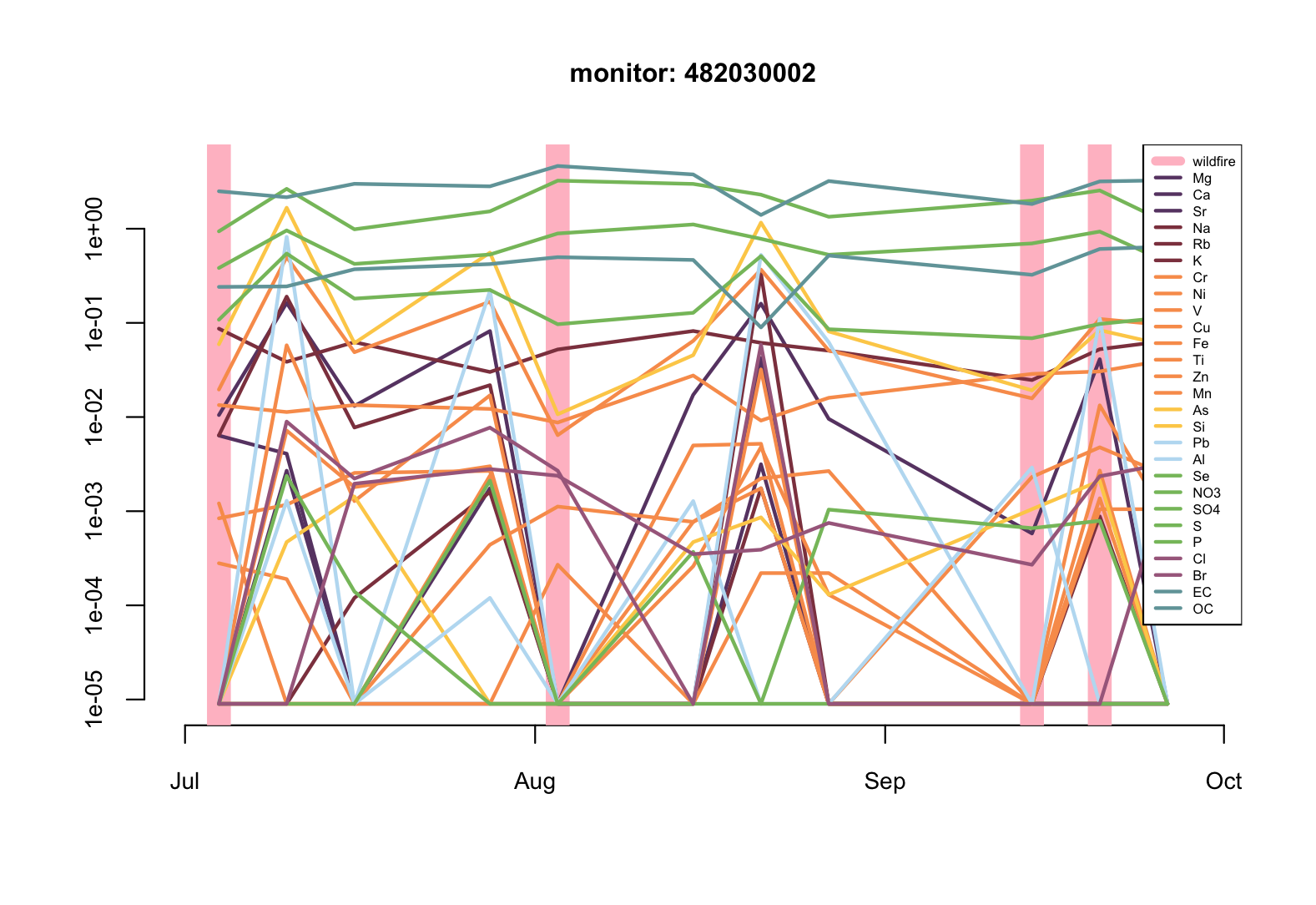} \;
\includegraphics[width=7cm]{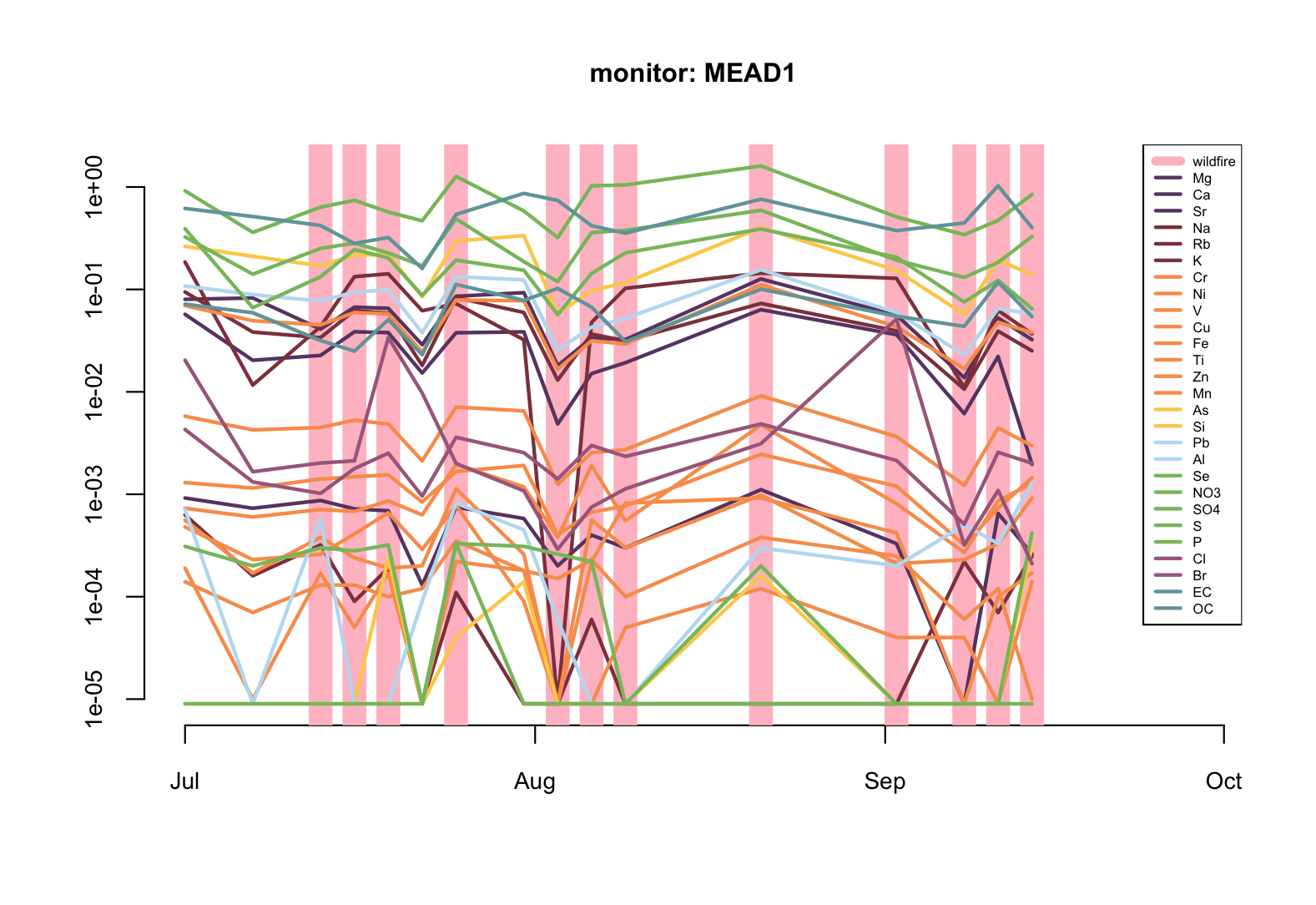} 
\caption{Representation of the time series of the observed 27 chemical species for six monitors in the consider period: summer 2014.The vertical pink lines indicate the presence of the wildfire smoke in that specific day. \label{fig:time_series}}
\end{stdfigure} 

\bigskip \section*{\small Definition of treatment: wildfire smoke day}

The concentration of wildfire smoke \PM is derived from the daily prediction of \citet{childs2022daily}, available from the GitHub repository \href{https://github.com/echolab-stanford/daily-10km-smokePM}{echolab-stanford/daily-10km-smokePM}. 
\review{Specifically, \citet{childs2022daily} identify ``smoke days'' using satellite plumes and air trajectories, estimate smoke-related \PM anomalies at Environmental Protection Agency (EPA) monitoring stations, and estimate a predictive model to account for spatial gaps and missing data over a daily $10\times 10 Km$ grid across the contiguous US.}

\review{For our application, we use the results of \citet{childs2022daily} to define the treatment variable ``smoke day''. Specifically, for each monitor-day unit, the treatment level $t = 1$ indicates a ``smoke day'' when a non-zero prediction of smoke-related \PM is predicted at that monitor location, and $t = 0$ for ``non-smoke day'' when no smoke-related \PM is predicted.} The \review{estimation} resolution in the grid is $10 Km$ (i.e., each unit aggregates the information over $10\times 10 Km$ area), therefore we match the location of each air pollution monitor with the closest point on the grid.

\bigskip \section*{\small Study Design}
Our analysis focuses on July–September 2014, when wildfire smoke is most prevalent in the United States \citep{krasovich2025influence}. The observational unit is a monitor-day. Since chemical concentrations are measured every three days, the dataset contains $7,467$ units. 
The multivariate outcome is the concentration of $27$ chemical species, measured in $\mu g/m^3$ and transformed on the natural logarithmic scale. Confounders include weather information, census data, type of monitor indication, month, latitude and longitude of monitor location, and type of monitor location (rural, city or intermediate).

Initially we have $5,754$ monitor day units assigned in the control group --- that is, no wildfire smoke day at that monitor location --- and $1,713$ in the treated group --- that is, exposed to wildfire smoke. 
However, we use matching before running our model to make our analyses as robust as possible with respect to the potential measured confounding bias, using confounders to measure the similarity of the units in the two groups. This is common in observational studies  \citep{rosenbaum1983central} and in research on air pollution effects on health \cite[see, e.g., ][]{lee2021discovering, zorzetto2024bayesian}.

We employ a 1-to-1 nearest neighbor propensity score matching, yielding a balanced sample of $3,426$ units, and improving the balance of the covariates. The reduction in units is due to the different sample sizes of the treated and control groups in the original data, and 1-to-1 matching creates a sample with the same size for the treated and control groups. The causal effects, estimated in the main text, target the sample obtained through the matching procedure.

\bigskip

\section*{References}
Albert, J. H. and Chib, S. (2001). Sequential ordinal modeling with applications to survival data. {\it Biometrics 57}, 829–836. \\
\indent Bhattacharya, A. and Dunson, D. B. (2011). Sparse bayesian infinite factor models. {\it Biometrika 98}, 291–306. \\
\indent Childs, M. L. et al. (2022). Daily local-level estimates of ambient wildfire smoke \PM for the contiguous us. {\it Environmental Science \& Technology 56}, 13607–13621.\\
\indent Krasovich Southworth, E. et al. (2025). The influence of wildfire smoke on ambient \PM chemical species concentrations in the contiguous us. {\it  Environmental Science \& Technology} . \\
\indent Lee, K., Small, D. S., and Dominici, F. (2021). Discovering heterogeneous exposure effects
using randomization inference in air pollution studies. {\it Journal of the American Statistical Association 116}, 569–580. \\
\indent Rodriguez, A. and Dunson, D. B. (2011). Nonparametric Bayesian models through probit stick-breaking processes. {\it Bayesian Analysis 6}, 1.\\
\indent Rosenbaum, P. R. and Rubin, D. B. (1983). The central role of the propensity score in observational studies for causal effects. {\it Biometrika 70}, 41–55. \\
\indent Zorzetto, D. et al. (2024). Bayesian nonparametrics for principal stratification with continuous post-treatment variables. {\it arXiv preprint arXiv:2405.17669}.

\end{document}